\documentclass[11pt,a4paper]{article}
 \pdfoutput=1
 \usepackage{jheppub}
 \usepackage{slashbox}
 \usepackage{mathrsfs}
 \usepackage{bbold}
 \usepackage{cancel}

\def\cF{{\cal F}} \def\cG{{\cal G}}

\def\cL{{\cal L}} 
 \def\cO{{\mathcal O}}
 
\def\cR{{\cal R}}

\newcommand{\hc}{\text{h.c.}}
\newcommand{\unity}{\mathbb{1}}

\newcommand{\mean}[1]{\langle#1\rangle}

\newcommand{\ov}[1]{\overline{#1}}

\newcommand{\beq}{\begin{equation}}
\newcommand{\eeq}{\end{equation}}
\newcommand{\bac}{\begin{equation}\begin{array}}
\newcommand{\eac}{\end{array}\eeq}
\newcommand{\ba}{\begin{array}}
\newcommand{\ea}{\end{array}}
\newcommand{\bea}{\begin{eqnarray}}
\newcommand{\eea}{\end{eqnarray}}

\leftline{TUM-HEP-802/11}
\leftline{FLAVOUR(267104)-ERC-2}
\bigskip

 \title{\boldmath The Impact of Flavour Changing Neutral Gauge Bosons on $\ov{B}\to X_s\gamma$}

 \author[a,b]{Andrzej J. Buras,} 
 \author[a,b]{Luca Merlo,} 
 \author[a,b,c]{and Emmanuel Stamou} 

\affiliation[a]{{Physik-Department, Technische Universit\"at M\"unchen, \\
  	       James-Franck-Strasse, D-85748 Garching, Germany}}
\affiliation[b]{{TUM Institute for Advanced Study, Technische Universit\"at M\"unchen,\\
  	       Lichtenbergstrasse 2a, D-85748 Garching, Germany}}
\affiliation[c]{Excellence Cluster Universe, Technische Universit\"at M\"unchen,\\
  	       Boltzmannstrasse 2, D-85748 Garching, Germany \\ }

\emailAdd{andrzej.buras@ph.tum.de}
\emailAdd{luca.merlo@ph.tum.de}
\emailAdd{emmanuel.stamou@ph.tum.de}

\abstract{
The branching ratio of the rare decay $\ov{B}\to X_s\gamma$ provides
potentially strong constraints on models beyond the Standard Model.
Considering a general scenario with new heavy neutral gauge bosons, present in particular
in $Z'$ and gauge flavour models, we point out two new contributions to
the $\ov{B}\to X_s\gamma$ decay. The first one originates from one-loop diagrams mediated by gauge bosons and heavy exotic quarks with electric charge $-1/3$. The second contribution stems from the QCD mixing of neutral current-current operators generated by heavy neutral gauge bosons and the dipole operators responsible for the $\ov{B}\to X_s\gamma$ decay. The latter mixing is calculated here for the 
first time.
We discuss general sum rules which have to be satisfied in any
model of this type. We emphasise that the 
neutral gauge bosons in question could also significantly affect other fermion radiative decays as well as non-leptonic two-body $B$ decays, $\epsilon'/\epsilon$, anomalous $(g-2)_\mu$ and electric dipole moments.}

\keywords{Rare Decays, Beyond Standard Model, Gauge Flavour Symmetries, $Z'$ models}
\arxivnumber{1105.5146}

\begin{document} 

\maketitle

\flushbottom


%
%
\section{Introduction}
\label{sec:Intro}

Extensions of the Standard Model (SM) with additional gauge symmetries are of particular interest in view of current direct and indirect
searches for physics beyond the SM. A feature of such theories is the existence of new gauge bosons.
These could provide clear deviations from the SM predictions and have been studied as possible early discoveries
at the LHC.

Constructions with a single additional $U(1)$ factor correspond to an extra neutral gauge boson $Z'$ \cite{Langacker:ZpReview, SVZ:Zp, SSVZ:Zp,ABFKS:Zp} and have received a lot of attention, since they are naturally present in a large variety of models, such as Grand Unified Theories
(GUTs), higher-dimensional models, little or composite Higgs models, superstring constructions.
Hence, the predicted $Z'$ mass spans a wide range of values, from the high GUT scale down to
the TeV scale, having different phenomenological consequences. On the other hand,
it is difficult to determine model independent bounds on the $Z'$ mass, since its relation to
observables depends strongly on the mixing of the $Z'$ to the SM $Z$, the strength of the new gauge-coupling constant, the charges of the SM fermions and Higgs doublets, the presence of additional exotic
fermions necessary to cancel the anomalies, etc..

Still, there exist bounds on the $Z'$ mass and its mixing with the SM $Z$ from precision electroweak data, direct and
indirect searches at the Tevatron, and interference effects at LEP2 in specific $Z'$ models
(see~\cite{Langacker:ZpReview} and references therein). Assuming an electroweak strength for the new gauge coupling constant and family-universal SM fermion charges, most of the canonical $Z'$ models have predictions in agreement with the experimental
constraints, for $Z'$ masses not too much below $1$ TeV. However,
considering leptophobic $Z'$ models or models with family non-universal SM fermion charges, the $Z'$
mass is allowed to be even smaller.

Beyond the simple $U(1)$ group, extended gauge symmetries have also attracted a lot of attention, especially to explain the observed pattern of fermion masses and mixings (some of the earliest papers on this subject are~\cite{Mohapatra:GaugedSym,BZ:GaugedSym, Ong:GaugeSym, WZ:GaugeSym, Chakrabarti:GaugeSym, MY:GaugeSym, DKW:GaugeSym, CPM:GaugeSym, BC:GaugeSym, Berezhiani:GaugeSym, BK:GaugeSym}). A common feature of these models is the appearance of dangerous flavour-changing-neutral-current (FCNC) contributions, mediated by the new flavour gauge bosons. Hence, a high new physics (NP) scale, usually larger than $1000$ TeV, must be adopted to provide the necessary suppressions. This prevents any NP direct observation at the present and future colliders.

An alternative approach to provide an explanation of the pattern of fermion masses is considering global flavour symmetries; in particular discrete symmetries have been widely studied in the last years (for a review see \cite{AF:ReviewDiscreteSyms}). It is interesting to notice that the FCNC contributions are usually less constraining in these models~\cite{IKOT:LFVA41, FHLM:LFVinA4, HIST:LFVA42, IKOST:LFVDelta54, FHM:Vacuum, FHLM:LFVinSUSYA4, HMP:LFVinSUSYA4SS, FP:RareDecaysA4, MRZ:FVTp, ABMP:Constraining1, ABMP:Constraining2} compared to gauge flavour models, opening the possibility for a lower NP scale and correspondingly for new particles beyond SM that can be detected in present and future colliders.

Recently, a series of papers \cite{GRV:SU3gauged, Feldmann:SU5gauged, GMS:SU3gaugedLR, AFM:EffectiveGaugeSym} appeared suggesting the possibility of explaining fermion masses and mixings with a NP scale of a few TeV, but with extra gauge symmetries. In these models, the gauge flavour symmetry is the product of non-Abelian $U(3)$ factors. For each gauge-group
generator there is a new gauge boson, with no color or electric charge. In addition, new exotic fermions have to be introduced
to cancel the anomalies of the new gauge sector.

The analyses in \cite{AFM:EffectiveGaugeSym,GRV:SU3gauged, Feldmann:SU5gauged, GMS:SU3gaugedLR} and most papers to date studied almost exclusively the impact of flavour changing heavy neutral gauge bosons on $\Delta F=2$ transitions and rare $B$ and $K$ decays, leaving their impact on the radiative decay $\ov{B}\to X_s\gamma$ aside. As the latter provides generally a very strong constraint on extensions of the SM, it is important to investigate whether the presence of neutral gauge bosons with flavour violating couplings and new exotic heavy fermions would put the models in question into difficulties and whether generally the role of such new gauge bosons and new fermions in the $\ov{B}\to X_s\gamma$ decay is significant. 

Having this in mind, we point out in this paper two new contributions to the $\ov{B}\to X_s\gamma$ decay, that to our knowledge
have not been considered in the literature so far. The first one originates from one-loop diagrams mediated by  
{\it heavy neutral} gauge bosons and {\it exotic} quarks with electric charge $-1/3$, when {\it flavour-violating left-handed and right-handed}
couplings of these gauge bosons to SM and exotic quarks are present. Then, this new physics contribution to the Wilson
coefficient of the dominant operator  is enhanced by $m'_F/m_b$ with $m_b$ being the mass of the $b$-quark and $m'_F$ the mass of the exotic fermion, in general significantly heavier than SM ones. 

Analogous enhancement is known from the study of left-right symmetric models \cite{CCFKM:BdecaysRLsym,AI:BSGdecayLRsym,CM:BSGdecayLRsym}, but the absence of new heavy fermions and the smallness of $W_L-W_R$ mixing makes this contribution,  mediated dominantly by {\it charged} SM gauge bosons with right-handed couplings, sufficiently small to agree with data. 

Within the littlest Higgs model with T-parity, similar contributions to the $\ov{B}\to X_s\gamma$ decay, mediated by heavy neutral gauge bosons and heavy mirror fermions, have already been considered  \cite{BBPTU:LittlestHiggsT}. However, in this context only left-handed currents are present and in addition the diagrams in question are strongly suppressed by the mixing angles in the T-odd sector and by the GIM mechanism, so that these contributions turn out to be very small.

The second new contribution stems solely from the heavy neutral gauge bosons, independently of the presence of exotic fermions in the theory. These gauge bosons are integrated out at the scale of the order of their masses, in general much higher than $\mu_b\approx m_b$. The resulting {\it neutral current-current} dimension-six operators affect the $\ov{B}\to X_s\gamma$ rate through QCD mixing with the magnetic dipole operator responsible for the $\ov{B}\to X_s\gamma$ decay. This contribution is present when 
the new neutral gauge bosons have direct couplings to the $b$ and $s$ quarks.

As we shall demonstrate explicitly, these new contributions imply constraints on the parameters of models,
containing neutral flavour gauge bosons with couplings described above and with masses within the reach of the LHC. We shall also identify the conditions under which the first of these contributions turns out to be negligible, such as very small mixings among SM and exotic quarks, naturally following from a fermion see-saw mechanism.

We point out that similar contributions could be considered in $f_j\to f_i \gamma$, like $\mu\to e\gamma$
and $t\to c\gamma$, where $f$ is any SM fermion, mediated by exotic fermions with the same electric charge as $f$. The same comments apply to flavour conserving observables like  $(g-2)_\mu$
and EDMs, in which dipole operators play the dominant role. Moreover, our QCD analysis of the mixing among neutral current-current and QCD-penguin operators is also applicable to other processes, such as non-leptonic two-body $B$ decays, and other observables, like $\epsilon'/\epsilon$.

We describe the general framework and fix the notation that will be used throughout the rest of the paper in Sect.~\ref{sec:Context}. In Sect.~\ref{sec:analytic} we present the analytical results for the exotic-fermion contribution
to the $\ov{B}\to X_s\gamma$ decay in the context of only one neutral flavour gauge boson and
one exotic quark with electric charge $-1/3$. Subsequently, we generalise our results to an arbitrary number of
such gauge bosons and quarks. 

In Sect.~\ref{sec:QCD} we discuss the second new contribution to the $\ov{B}\to X_s\gamma$ decay. We extend
the SM QCD renormalisation group (RG) analysis by taking into account all the new neutral current-current operators,
generated through the exchange of a single neutral gauge boson $A_H$. To this end we calculate the relevant anomalous dimension matrices at the leading order (LO).

In Sect.~\ref{sec:pheno} we present our model-independent results in a general form, that can be easily applied to all models with flavour-violating neutral gauge bosons, like $Z'$-models or models with non-Abelian gauge flavour symmetries. We identify three major classes of such models; for each we cast our results into simple formulae, that depend only on the masses and couplings of the new particles. Finally, we illustrate how these formulae can be translated into constraints on the parameter space of representative toy-models. In Sect.~\ref{sec:concl} we summarise and conclude.

%
%
\section{Notation and fermion see-saw mechanism}
\label{sec:Context}

We focus on models in which the new physics contributions to
$\ov{B}\to X_s\gamma$ discussed in the previous section are present. In what follows, we describe a general framework and we fix the notation.

We consider the general case in which a product of $k\geq1$ gauge flavour groups,
\mbox{$\cG_f=\cG_1\times\ldots \times\cG_k$}, is added to the SM gauge group. Depending
on the dimension $d_i$ of each symmetry factor, $d_1+\ldots+d_k$ new flavour
gauge bosons enrich the particle spectrum. In general, the SM fermions $\psi$ transform
under $\cG_f$ and consequently anomalies from gauge or mixed gauge-gravitational triangles
can arise. In this case, it is possible to introduce a new set of exotic fermions $\Psi$,
suitably transforming under the SM and gauge flavour groups to obtain an anomaly-free theory. 

On the other hand, the presence of these new exotic fermions is a quite general feature of
flavour models, when considering renormalisable ultraviolet completions. This holds not
only for gauge flavour symmetries, that we are discussing in this paper, but also for global
flavour symmetries. Indeed, in these models the new exotic fermions and their mixing with
the SM ones account for the mass spectrum of the SM fermions (see \cite{BGPZ:HeavyFermions} and references therein).
This typically happens through a fermion see-saw mechanism where the small masses of
the SM fermions are explained by the large masses of their corresponding exotic fermions.
Without loss of generality, we consider only the case of $SU(2)_L$-singlet exotic
fermions. Hence the see-saw Lagrangian can be schematically written as
\beq
  \cL_{\text{see-saw}}= \ov{\Psi}_L\, M\, \Psi_R + \ov{\psi}_L\, Y^D_1\, \Psi_R\, \phi + \ov{\Psi}_L\, M^D_2\, \psi_R+\hc\,,
\label{SeesawLag}
\end{equation}
where $\psi_{L,R}$ ($\Psi_{R,L}$) comprise all left- and right-handed SM
(exotic) fermion fields; their dimensions also define the dimensions
of the matrices $M$, $Y^D_1$ and $M^D_2$. In the same expression,
$\phi$ represents the SM Higgs field.

This see-saw Lagrangian can be made invariant under $\cG_f$ if both SM and exotic
fermions belong to the same representation of $\cG_f$ and $M$, $Y^D_1$, and $M^D_2$ are proportional
to the identity matrix in the flavour space. This case, however, is phenomenologically not interesting,
because then the flavour symmetry is exactly preserved and it is not possible to reproduce
the observed pattern of SM fermion masses and mixings. 

On the other hand, a more realistic situation  appears when we impose the invariance under
$\cG_f$ at high scales, by introducing additional degrees of freedom, and then break the flavour symmetries at lower scales to provide the correct SM fermion masses and mixings. In the see-saw Lagrangian, the invariance under $\cG_f$ and its breaking mechanism can be achieved in
various ways: we can promote $M$, $Y_1^D$, and $M_2^D$, or only some of them, either to spurion fields 
\`a la MFV \cite{DGIS:MFV,CGIW:MLFV,DP:MLFV,AIMMN:VMLFV}, or to dynamical scalar fields with non-vanishing
vacuum expectation values (VEVs) \cite{FN}, or to fermion condensates \cite{CG:MFV}. In what follows, we
simply treat the terms with $M$, $Y_1^D$, and $M_2^D$, or only some of them, as flavour violating
terms originating from some non-specified dynamics at the high scale \cite{AGMR:ScalarPotentialMFV}.

In the simplest phenomenologically interesting scenario, we discuss, the fields $\psi_{L}$ ($\psi_{R}$) and $\Psi_{R}$ ($\Psi_{L}$) transform
under the same representation of a single flavour group $\cG_f$. The see-saw Lagrangian is the
same as in Eq.~(\ref{SeesawLag}),  with the only flavour symmetry breaking parameter being
$M$, while $Y^D_1$ and $M^D_2$ are family universal parameters. In general, the square matrix $M$ has
 non-diagonal hierarchical entries in order to provide the
correct description of the SM fermion masses and mixings. Before finding
the mass eigenvalues of the SM and exotic fermions, it is useful
to diagonalise $M$. In general, $M$ is diagonalised by a bi-unitary transformation
\begin{equation}
\hat{M}=V^\dag_L\, M\, V_R\,,
\end{equation}
where $V_{L,R}$ are unitary matrices and $\hat{M}$ is a diagonal matrix. After the absorption of $V_L$ ($V_R$) in $\Psi_L$ ($\Psi_R$) and $\psi_R$ ($\psi_L$) by a field redefinition and the breaking of the electroweak symmetry, the resulting mass matrix is as follows:
\beq
\left(\ov{\psi}_L,\, \ov{\Psi}_L\right)
\left(
        \begin{array}{cc}
             0					& Y_1^D\mean{\phi} \times\unity\\[2mm]
	    M_2^D\times\unity 		& \hat{M}\\
        \end{array}
\right)
\left(
        \begin{array}{c}
             \psi_R\\
	    \Psi_R \\
        \end{array}
\right)\,.
\eeq

Reabsorbing possible unphysical phases, we can now define the mass eigenstates $f_{L,R}$ and $F_{L,R}$ by performing orthogonal rotations
on the flavour eigenstates $\psi_{L,R}$ and $\Psi_{L,R}$. For each generation we may write
\begin{equation}
\left(
         \begin{array}{c}
             f_{L,R}  \\
             F_{L,R}   \\
         \end{array}
\right)=
\left(
         \begin{array}{cr}
            \cos\theta_{L,R}  & -\sin\theta_{L,R}  \\
            \sin\theta_{L,R}  & \cos\theta_{L,R} \\
         \end{array}
\right)\left(
         \begin{array}{c}
              \psi_{L,R}  \\
              \Psi_{L,R}  \\
         \end{array}
\right)\,.
\end{equation} 
Denoting with $m_{f_k}$ the mass of the light fermion $f_k$ and with $m'_{F_k}$ that of
the heavy exotic fermion $F_k$ and using for brevity the notation $M_1^D\equiv Y_1^D \mean{\phi}$,
what follows is a direct inverse proportionality between $m_{f_k}$ and $m'_{F_k}$, as in the usual see-saw mechanism:
\begin{equation}
m_{f_k}m'_{F_k}=M_1^D\,M_2^D\,.
\eeq
We can express such masses in terms of the flavour symmetry breaking terms:
\beq
m_{f_k}=\dfrac{\sin\theta_{R_k}\,\sin\theta_{L_k}}{\cos^2\theta_{R_k}-\sin^2\theta_{L_k}}\hat{M}_k\,,\qquad\qquad
m'_{F_k}=\dfrac{\cos\theta_{R_k}\,\cos\theta_{L_k}}{\cos^2\theta_{R_k}-\sin^2\theta_{L_k}}\hat{M}_k\,,
\end{equation}
where a straightforward calculation gives
\begin{equation}
\sin\theta_{L_k}=\sqrt{\dfrac{m_{f_k}}{M_2^D}\dfrac{M_1^D\,m'_{F_k}-M_2^D m_{f_k}}{m^{\prime\,2}_{F_k}-m_{f_k}^2}}\,,\qquad\quad
\sin\theta_{R_k}=\sqrt{\dfrac{m_{f_k}}{M_1^D}\dfrac{M_2^D\,m'_{F_k}-M_1^D m_{f_k}}{m^{\prime\,2}_{F_k}-m_{f_k}^2}}\,.
\label{FormulaSin1}
\end{equation}

These results are exact and valid for all the fermion generations. However, taking
the limit in which $m'_{F_k}\gg m_{f_k}$ and $M_1^D\approx M_2^D\equiv M^D$,
we find simple formulae that transparently expose the behaviour of the previous expressions.
In this limit we find
\begin{equation}
m_{f_k}\approx \dfrac{(M^D)^2}{\hat{M}_k}\,,\qquad\qquad
m'_{F_k}\approx \hat{M}_k\,,\qquad\qquad
\sin\theta_{L_k}\approx \sin\theta_{R_k} \approx\sqrt{\dfrac{m_{f_k}}{m'_{F_k}}}\,,
\label{FormulaSin2}
\end{equation}
as in the usual see-saw scheme in the limit $\hat{M}_k\gg M_{1,2}^D$. Notice that these
simplified relations are valid for all the fermions, apart from the top quark for which the condition $\hat{M}_3\gg M^D_{1,2}$ is not satisfied and large corrections  to Eq.~(\ref{FormulaSin2}) are expected.

When the flavour symmetry is broken, the flavour gauge bosons develop masses proportional to the flavour breaking parameters. It is not possible to enter into details without specifying a particular model and in the following we simply assume that the neutral gauge boson masses are controlled by the parameters $\hat{M}_k$, as for the exotic fermions.

The part of the Lagrangian containing the kinetic terms and the usual couplings among
fermion and gauge bosons is given by 
\begin{equation}
\cL_{kin}=i\,\left(\ov{\psi}\, D\hspace{-2.5mm}\big/\, \psi +\ov{\Psi}\, D\hspace{-2.5mm}\big/\,\Psi\right)\,.
\label{StandardKineticTerms}
\end{equation}
Since we are only interested in discussing the couplings between fermions and the new gauge bosons, in writing the relevant covariant derivative we do not consider the coupling terms with the SM gauge bosons but only the following term
\begin{equation}
D^\mu\supset i\, \dfrac{g_H}{2} \sum_{a}A_{H_a}^\mu\, T_\cR^a\,.
\end{equation}
In this expression, $g_H$ is the flavour-gauge coupling, $T_\cR^a$ are the generators of the flavour
group $\cG_f$ in the representation $\cR$, and $A_{H_a}$ the corresponding
flavour gauge bosons. The resulting Feynman rules for a single gauge boson $A_H$ in the fermion-mass eigenbasis, $\cF\equiv\{F,\,f\}$, are given by the following relations:
\begin{equation}
\includegraphics{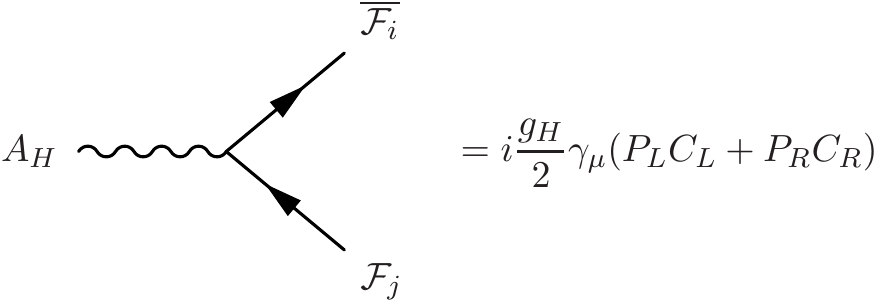}
\label{vertexGeneral}
\end{equation}
with
\begin{equation}
\label{Couplingff}
A_H\,\ov{f}_i\, f_j:\begin{cases}
C_L=\cos\theta_{L_i}\left(V_R^\dag T_\cR^a V_R\right)_{ij}\cos\theta_{L_j}+\sin\theta_{L_i}\left(V_L^\dag T_\cR^a V_L\right)_{ij}\sin\theta_{L_j}\\
C_R=\cos\theta_{R_i}\left(V_L^\dag T_\cR^a V_L\right)_{ij}\cos\theta_{R_j}+\sin\theta_{R_i}\left(V_R^\dag T_\cR^a V_R\right)_{ij}\sin\theta_{R_j}
\end{cases}
\end{equation}
\begin{equation}
\label{CouplingFF}
A_H\,\ov{F}_i\, F_j:\begin{cases}
C_L=\sin\theta_{L_i}\left(V_R^\dag T_\cR^a V_R\right)_{ij}\sin\theta_{L_j}+\cos\theta_{L_i}\left(V_L^\dag T_\cR^a V_L\right)_{ij}\cos\theta_{L_j}\\
C_R=\sin\theta_{R_i}\left(V_L^\dag T_\cR^a V_L\right)_{ij}\sin\theta_{R_j}+\cos\theta_{R_i}\left(V_R^\dag T_\cR^a V_R\right)_{ij}\cos\theta_{R_j}
\end{cases}
\end{equation}
\begin{equation}
\label{CouplingfF}
A_H\,\ov{f}_i\, F_j:\begin{cases}
C_L=\cos\theta_{L_i}\left(V_R^\dag T_\cR^a V_R\right)_{ij}\sin\theta_{L_j}-\sin\theta_{L_i}\left(V_L^\dag T_\cR^a V_L\right)_{ij}\cos\theta_{L_j}\\
C_R=\cos\theta_{R_i}\left(V_L^\dag T_\cR^a V_L\right)_{ij}\sin\theta_{R_j}-\sin\theta_{R_i}\left(V_R^\dag T_\cR^a V_R\right)_{ij}\cos\theta_{R_j}
\end{cases}
\end{equation}
\begin{equation}
\label{CouplingFf}
A_H\,\ov{F}_i\, f_j:\begin{cases}
C_L=\sin\theta_{L_i}\left(V_R^\dag T_\cR^a V_R\right)_{ij}\cos\theta_{L_j}-\cos\theta_{L_i}\left(V_L^\dag T_\cR^a V_L\right)_{ij}\sin\theta_{L_j}\\
C_R=\sin\theta_{R_i}\left(V_L^\dag T_\cR^a V_L\right)_{ij}\cos\theta_{R_j}-\cos\theta_{R_i}\left(V_R^\dag T_\cR^a V_R\right)_{ij}\sin\theta_{R_j}
\end{cases}
\end{equation}

A few comments are in order:
\begin{description}
\item 1)\quad These results are valid for both Abelian and non-Abelian flavour gauge
  symmetries. In particular, barring accidental cancellations,
  the couplings described in Eqs.~(\ref{CouplingfF}) and (\ref{CouplingFf}) are always non-vanishing. The only exception consists in all those $Z'$ models in which the SM fermions have universal charges. In these cases the flavour gauge symmetry $\cG_f$ is a simple $U(1)$ and the charge universality assures that $T_\cR^a\propto\unity$. Hence, the couplings in Eqs.~(\ref{CouplingfF}) and (\ref{CouplingFf}) identically vanish and the couplings in Eqs.~(\ref{Couplingff}) and (\ref{CouplingFF}) turn out to be flavour conserving.

\item 2)\quad When the flavour symmetry is a product of different groups, \mbox{$\cG_f=\cG_1\times\ldots \times \cG_k$}, and fermions transform non-trivially under different factors of $\cG_f$, Eqs.~(\ref{Couplingff})--(\ref{CouplingFf})
  change. In particular, if $\psi_L$ ($\Psi_L$) and $\psi_R$ ($\Psi_R$) transform under
  two distinct groups, then only one term in the right-hand side of the previous
  equations is present (for example in Eq.~(\ref{Couplingff}), $C_L$ could have either the term \mbox{$\cos\theta_{L_i}\left(V_R^\dag T_\cR^a V_R\right)_{ij}\cos\theta_{L_j}$} or \mbox{$\sin\theta_{L_i}\left(V_L^\dag T_\cR^a V_L\right)_{ij}\sin\theta_{L_j}$}, but not both terms simultaneously). As a result, even for the simple case $\cG_f=U(1)^2$ with
  universal fermion charges, the couplings described in Eqs.~(\ref{CouplingfF}) and
  (\ref{CouplingFf}) are non-vanishing. Notice that in this particular case all the couplings in Eqs.~(\ref{Couplingff})--(\ref{CouplingFf}) turn out to be flavour conserving.
  
\item 3)\quad When the SM fermion masses originate from the see-saw mechanism with heavy exotic fermions, $\sin\theta_{L_k}$ and $\sin\theta_{R_k}$ are small (cf. Eqs.~(\ref{FormulaSin1}) and (\ref{FormulaSin2})). Hence, the couplings described in Eqs.~(\ref{CouplingfF})  and (\ref{CouplingFf}) are all suppressed. In the particular case of $m'_{F_k}\gg m_{f_k}$, this suppression is roughly given by $\sqrt{m_{f_k}/m'_{F_k}}$.

\item 4)\quad When a theory is anomaly free without the necessity of introducing extra exotic fermions, only the expressions in Eq.~(\ref{Couplingff}) apply. Still the couplings can be flavour violating. This is the case of $U(1)$ models in which either the charge assignment of the SM fermions is such that no anomalies arise or the Green-Schwarz mechanism \cite{GS:GSmechanism} is implemented into the theory to compensate for the anomalies. In the specific case of $Z'$ models with universal fermion $U(1)$-charges, the couplings are flavour conserving, as already discussed in 1).
\end{description}

As we will show in Sects.~\ref{sec:analytic} and \ref{sec:QCD}, in virtually any gauge model, in which SM fermion masses are explained through a see-saw mechanism with heavy exotic fermions, both new contributions we introduced in Sect.~\ref{sec:Intro} are present and have potentially interesting effects. Their impact depends on the details of the specific model, according to the above discussion.

On the other hand, for completeness we will also consider models with gauge symmetries beyond those of the SM that provide either an explanation of the fermion mass patterns through a different mechanism than the see-saw or no explanation at all. In these cases, heavy exotic fermions might not be present in the particle spectrum, as in 4), and the corresponding contributions are absent. However, when present, the couplings listed in Eqs.~(\ref{Couplingff})--(\ref{CouplingFf}) have in general a completely different structure. In particular there might be no suppressions of the exotic fermion contributions due to $\sin\theta_{L_k}$ and $\sin\theta_{R_k}$.

The presence of exotic fermions can modify the couplings
of the SM W and Z bosons to fermions with respect to their couplings within the
SM. As a consequence, non-unitary quark and lepton mixing matrices and
flavour violating Z couplings
are usual results of such changes, which are specific to the considered gauge
flavour model. Also these modifications affect only the Wilson coefficients and do not generate further operators. For this reasons we shall not perform
such a study in our model-independent analysis. A study in a model-dependent
context considering both SM and heavy gauge bosons contributing to the
$\ov{B}\to X_s\gamma$ branching ratio is in progress \cite{BCMS:progress}.

%
%
\section{New contributions at the high scale}
\label{sec:analytic}
\mathversion{bold}
\subsection{Effective Hamiltonian for $b\to s\gamma$}
\mathversion{normal}
\label{subsec:ContributionGeneralStructure}

Adopting the overall normalisation of the SM effective Hamiltonian and considering 
as a first step only the dipole operators and the contribution of new
physics to their Wilson coefficients, we write the effective Hamiltonian
relevant for $b\to s\gamma$ at the high matching scale $\mu_H\gg M_W$ as: 
\begin{equation}
\label{Heff_at_mu}
{\cal H}_{\rm eff}^{(b\to s\gamma)} = - \dfrac{4 G_{\rm F}}{\sqrt{2}} V_{ts}^* V_{tb} \Big[\Delta C_{7\gamma}(\mu_H) Q_{7\gamma} + \Delta  C_{8G}(\mu_H) Q_{8G} +\Delta  C'_{7\gamma}(\mu_H) Q'_{7\gamma} +\Delta  C'_{8G}(\mu_H) Q'_{8G} \Big]\,.
\end{equation}
At this scale, the $W$ bosons are still dynamical and the usual SM contribution
to the Wilson coefficients is absent. We will include it after integrating out the $W$
boson at the lower scale $\mu_W\approx \cO(M_W)$. In our conventions the dipole operators are given by:
\begin{equation}
\label{O6B}
\begin{aligned}
Q_{7\gamma}  &=  \dfrac{e}{16\pi^2}\, m_b\, \bar{s}_\alpha\, \sigma^{\mu\nu}\, P_R\, b_\alpha\, F_{\mu\nu}\,,\\[2mm]          
Q_{8G}     &=  \dfrac{g_s}{16\pi^2}\, m_b\, \bar{s}_\alpha\, \sigma^{\mu\nu}\, P_R\, T^a_{\alpha\beta}\, b_\beta\, G^a_{\mu\nu}  
\end{aligned}
\end{equation}
and the primed operators are obtained from them after replacing the right-handed projector,
$P_{R}$, with the left-handed one, $P_L$. In the SM the contributions of the primed
operators are suppressed by $m_s/m_b$ relative to those coming from $Q_{7\gamma}$ and
$Q_{8G}$. We decompose the Wilson coefficients at the scale $\mu_W$
as the sum of the SM contribution and the new one from the exchange of neutral
flavour gauge bosons, after evolving the new physics contribution using the
RG running of the operators $Q^{(\prime)}_{7\gamma}$ and
$Q^{(\prime)}_{8G}$, namely
\begin{equation}
C_i(\mu_W)=C_i^{SM}(\mu_W)+\Delta C_i(\mu_W)
\end{equation}
and similarly for the primed coefficients. The SM Wilson coefficients
at LO are \cite{IL:Loops}
\begin{align}
\label{c7}
C^{SM}_{7\gamma} (\mu_W) &= \frac{3 x_t^3-2 x_t^2}{4(x_t-1)^4}\ln x_t - \frac{8 x_t^3 + 5 x_t^2 - 7 x_t}{24(x_t-1)^3}\,,\\[2mm]
\label{c8}
C^{SM}_{8G}(\mu_W) &= \frac{-3 x_t^2}{4(x_t-1)^4}\ln x_t - \frac{x_t^3 - 5 x_t^2 - 2 x_t}{8(x_t-1)^3}\,,
\end{align}
where $x_t\equiv m_t^2/M_W^2$. The corresponding primed coefficients are also given by
the previous expressions, but with an extra suppression factor of
$m_s/m_b$.

The new physics contributions at $\mu_W$, $\Delta C_i(\mu_W)$, follow from the RG evolution of the Wilson coefficients at $\mu_H$, $\Delta C_i(\mu_H)$, down to $\mu_W$; Their relations are discussed in Sect.~\ref{sec:QCD}. We can further decompose the new physics contributions at the high scale as the sum of two pieces, one coming from the exchanges of heavy exotic quarks and the second from the exchanges of light SM down-quarks: respectively,
\beq
\begin{aligned}
\Delta C_{7\gamma}(\mu_H)&=\Delta C^\text{heavy}_{7\gamma}(\mu_H) + \Delta C^\text{light}_{7\gamma}(\mu_H)\,,\\[2mm]
\Delta C_{8G}(\mu_H)     &=\Delta C^\text{heavy}_{8G}(\mu_H) + \Delta C^\text{light}_{8G}(\mu_H)\,.
\end{aligned}
\label{eq:wilsonatmhTot}
\eeq
In the following we analyse these two contributions separately.

\subsection{Contributions from heavy exotic quark exchanges}

\begin{figure}[h!]
\centering
\includegraphics{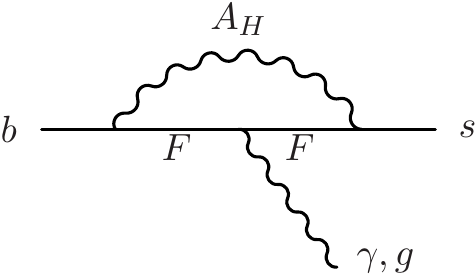}
\caption{\it Magnetic penguin diagram contributing to $b\to s\gamma$ and 
$b\to s g$ with internal $A_H$ and $F$ exchanges.
\label{fig:bsgamma}}
\end{figure}

We begin with the new contributions to $b\to s \gamma$ arising from the
exchange of heavy neutral flavour gauge bosons and heavy exotic quarks. In what follows,
we restrict the discussion to the contribution from a single virtual gauge
boson $A_H$ and a single virtual heavy quark of electric charge $-1/3$ and mass
$m'_F$. The results will then be generalised to an arbitrary number of such particles. 
From the general Feynman rules listed in Eqs.~(\ref{CouplingfF}) and (\ref{CouplingFf}),
we read off the flavour changing couplings of $A_H$ to the bottom and strange quarks and
the exotic quark $F$ relevant for $b\to s \gamma$:
\begin{equation}
\includegraphics{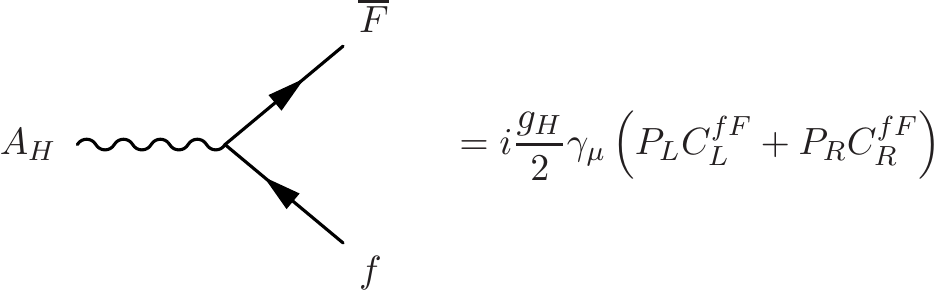}
\label{vertex}
\end{equation}
The only diagram \footnote{The corrections on the external legs, necessary for a canonical
kinetic term, are understood to be contained in finite off-diagonal field renormalisations.}
with a virtual $F$ and $A_H$ exchange contributing to the $b\to s\gamma$ transition in the
unitary gauge is shown in Fig.~\ref{fig:bsgamma}. In the same figure, we also show the
analogous diagram contributing to $b\to s\, g$. The latter will also contribute to the 
$\ov{B}\to X_s\gamma$ rate via QCD mixing of $Q_{8G}$ in $Q_{7\gamma}$. In both diagrams,
there is no triple gauge boson vertex as the heavy gauge boson $A_H$ considered does
not carry electric or colour charge.

We further decompose the Wilson coefficients $\Delta C^\text{heavy}_i$ at the high
scale as the sum of the SM-like LL contribution and a new LR one, where $L$ ($R$) stands for the $P_L$ ($P_R$) projector in the vertex of Eq.~(\ref{vertex}) involving the $s$($b$)-quark:
\beq
\begin{aligned}
\Delta C^\text{heavy}_{7\gamma}(\mu_H)&=\Delta^{LL} C^\text{heavy}_{7\gamma}(\mu_H) +\Delta^{LR}C^\text{heavy}_{7\gamma}(\mu_H)\,,\\[2mm]
\Delta C^\text{heavy}_{8G}(\mu_H)     &=\Delta^{LL} C^\text{heavy}_{8G}(\mu_H) +\Delta^{LR}C^\text{heavy}_{8G}(\mu_H)\,.
\end{aligned}
\label{eq:wilsonatmh}
\eeq
The corresponding primed coefficients to Eq.~(\ref{eq:wilsonatmh}) are obtained by interchanging $L\leftrightarrow R$ and including the suppression factor $m_s/m_b$ in the primed Wilson coefficients.

Following \cite{BBPTU:LittlestHiggsT}, the SM-like contributions LL and RR are
obtained by noting that their topological structure is the same as the SM {\it gluon}
magnetic penguin. Consequently, the known SM results for the Wilson coefficient of the
gluon dipole operator, $C^{SM}_{8G}$, can be used to obtain both LL and RR contributions
after a suitable redefinition of couplings and masses. Notice that in the neutral
exchange there is generally no analogue of a GIM mechanism, therefore mass independent terms
in the SM Wilson coefficients must be kept. We extract them from \cite{BMU:BSG1,BMU:BSG2}.
Our results are as follows:
\begin{equation}
\begin{aligned}
\Delta^{LL}C^\text{heavy}_{7\gamma}(\mu_H) &=-\dfrac{1}{6}\,\dfrac{g_H^2}{g_2^2}\,\dfrac{M_W^2}{M_{A_H}^2}\,\dfrac{C_L^{sF*}\,C_L^{bF}}{V_{ts}^*\,V_{tb}}\,\left(C_{8G}^{SM}(x)+\dfrac{1}{3}\right),\\[2mm]
\Delta^{LL}C^\text{heavy}_{8 G}(\mu_H) &= - 3 \Delta^{LL}C^\text{heavy}_{7\gamma}(\mu_H)\,,
\end{aligned}
\label{LLnew}
\end{equation}
with $x=m_F^2/M_{A_H}^2$. The $RR$ contributions are obtained by the
interchange $L\leftrightarrow R$. $C_{8G}^{SM}(x)$ is the SM function
on the right-hand side of Eq.~(\ref{c8}).\\

We now consider the LR and RL contributions, which have no SM equivalent. To obtain the
Wilson coefficients, we adapt to our case known calculations of $b\to sg$ in the context
of $b\to s\gamma$ in the left-right symmetric models \cite{CM:BSGdecayLRsym,BMU:BSG1},
where, however, the process is mediated by charged gauge bosons. In this way,
we obtain the following LR Wilson coefficients:
\begin{equation}
\begin{aligned}
\Delta^{LR}C^\text{heavy}_{7\gamma}(\mu_H)&=-\dfrac{1}{6}\,\dfrac{g_H^2}{g_2^2}\,\dfrac{M_W^2}{ M_{A_H}^2}\,\dfrac{m'_F}{m_b}\,\dfrac{C_L^{sF*}\,C_R^{bF}}{V_{ts}^*\,V_{tb}}\,C^{LR}_{8G}(x)\,,\\[2mm]
\Delta^{LR}C^\text{heavy}_{8G}(\mu_H)&= -3\Delta^{LR}C^\text{heavy}_{7\gamma}(\mu_H)\,,
\end{aligned}
\label{LRnew}
\end{equation}
with 
\beq
C^{LR}_{8G}(x)=\dfrac{-3x}{2(1-x)^3}\ln{x}+ \dfrac{3 x (x-3)}{4(x-1)^2} -1\,.
\label{CLR8Gcoeff}
\eeq
The analogous RL contributions are obtained by interchanging $L\leftrightarrow R$. 
In this case $C^{RL}_{8G}(x)=C^{LR}_{8G}(x)$.

The following properties should be noted:
\begin{description} 
\item 1)\quad the LR (RL) contribution in Eq.~(\ref{LRnew}) dominates over the LL (RR) one of Eq.~(\ref{LLnew}) in a large part of the parameter space, due to the factor $m'_F/m_b$; 
\item 2)\quad the factor $m'_F/m_b$ is replaced by $m_t/m_b$ in the usual left-right symmetric models, therefore in these models the enhancement is less pronounced; 
\item 3)\quad $C^{LR}_{8G}(x)$ is a non-vanishing monotonic function of $x$ and takes values in the range $[-1,\,-1/4]$ for $x$ from $0$ to $\infty$; 
\item 4)\quad in the decoupling limit these contributions turn out to vanish. Indeed when large masses for the exotic fermions $m'_{F_k}$ are considered, the masses of the neutral gauge bosons $M_{A_H}$ are also large, as we have already commented in Sect.~\ref{sec:Context}. Therefore the LL (RR) contributions approach zero with $1/M_{A_H}^2$ and the LR (RL) ones are strongly suppressed by $m'_F/M_{A_H}^2$.
\end{description}

\subsection{Contributions from light SM quark exchanges}

We now discuss $\Delta C^{\text{light}}_{7\gamma}$ and $\Delta C^{\text{light}}_{8G}$ from
Eq.~\eqref{eq:wilsonatmhTot}, namely the contributions arising from  the exchange of light down-type quarks in 
loop diagrams analogous to the one in Fig~\ref{fig:bsgamma}, where the flavour changing couplings of $A_H$ to the SM quarks read
\begin{equation}
\includegraphics{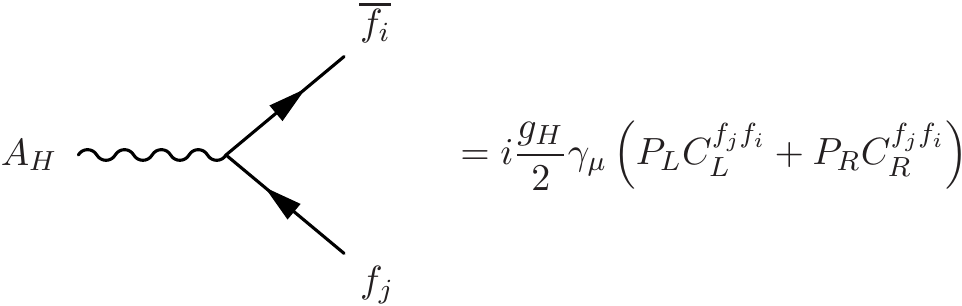}
\label{vertexLight}
\end{equation}
arising from the general Feynman rules listed in Eq.~(\ref{Couplingff}). In the SM, the light-quark contributions cancel each other at scales well above their masses due to the GIM mechanism. This is in contrast to the present case, where such a mechanism is absent.

Following the procedure of the previous section, we decompose the Wilson coefficients at the high scale as the sum of a $LL$ part and a $LR$ part:
\beq
\begin{aligned}
\Delta C^\text{light}_{7\gamma}(\mu_H)&=\Delta^{LL} C^\text{light}_{7\gamma}(\mu_H) +\Delta^{LR}C^\text{light}_{7\gamma}(\mu_H)\,,\\[2mm]
\Delta C^\text{light}_{8G}(\mu_H)     &=\Delta^{LL} C^\text{light}_{8G}(\mu_H) +\Delta^{LR}C^\text{light}_{8G}(\mu_H)\,.
\end{aligned}
\label{eq:wilsonatmhLight}
\eeq
In the framework of effective field theories the light degrees of freedom are treated as massless at the high matching-scale $\mu_H$.
Therefore, we only need to account for the first term in the expansion of the
light masses and only mass-independent terms in
Eq.~(\ref{LLnew}) and (\ref{LRnew}) contribute. For the LL SM-like contributions we find
\begin{equation}
\begin{aligned}
  \Delta^{LL}C^\text{light}_{7\gamma}(\mu_H) &=-\dfrac{1}{6}\,\dfrac{g_H^2}{g_2^2}\,
  						\dfrac{M_W^2}{M_{A_H}^2}\,\,\,
						\sum_{f}\dfrac{C_L^{sf*}\,C_L^{bf}}{V_{ts}^*\,V_{tb}}\,
						\left(\dfrac{1}{3}\right),\\[2mm]
  \Delta^{LL}C^\text{light}_{8 G}(\mu_H) &= - 3 \Delta^{LL}C^\text{light}_{7\gamma}(\mu_H)\,,
\end{aligned}
\label{eq:LLnewLight}
\end{equation}
and for the LR contribution we have from Eqs.~(\ref{LRnew}) and (\ref{CLR8Gcoeff})
\begin{equation}
\begin{aligned}
\Delta^{LR}C^\text{light}_{7\gamma}(\mu_H)&=-\dfrac{1}{6}\,
						\dfrac{g_H^2}{g_2^2}\,
						\dfrac{M_W^2}{ M_{A_H}^2}\,\,\,
						\sum_f\dfrac{m_f}{m_b}
						\dfrac{C_L^{sf*}\,C_R^{bf}}{V_{ts}^*\,V_{tb}}\,
						\left( -1\right)\,,\\[2mm]
\Delta^{LR}C^\text{light}_{8G}(\mu_H)&= -3\Delta^{LR}C^\text{light}_{7\gamma}(\mu_H)\,,
\end{aligned}
\label{eq:LRnewLight}
\end{equation}
where $f$ stands for the SM down-type quarks. Analogously, we obtain also the results for the primed Wilson coefficients. The
LL and RR SM-like contributions have already been considered in \cite{LP:ZpNonUniversal}, while
the LR and RL are new. Notice that the latter come with a factor
of $m_f/m_b$ and hence, the effects from $d$- and $s$-quarks are suppressed with
respect to the $b$-quark contribution.

\subsection{Generalisations and the GIM-like mechanism}

It is easy to extend the previous results to the case of an arbitrary number of neutral
flavour gauge bosons, $A_{H_i}$, and heavy fermions, $F_k$, by performing the
following substitution in Eqs.~(\ref{LLnew}) and (\ref{LRnew}):
\begin{equation}
\begin{aligned}
&\dfrac{g^2_H}{ M_{A_H}^2}\,C_{L,R}^{sF*}\,C_{L,R}^{bF}\,\left(C_{8G}^{SM}(x)+\dfrac{1}{3}\right)
\longrightarrow 
\sum_{i,k}\dfrac{g^2_{H_i}}{M_{A_{H_i}}^2}\,C_{L,R}^{ski*}\,C_{L,R}^{bki}\,\left(C_{8G}^{SM}(x_{ki})+\dfrac{1}{3}\right)\,,\\[2mm]
&g^2_H\,\dfrac{{m'_F}}{ M_{A_H}^2}\,C_{L,R}^{sF*}\,C_{R,L}^{bF}\,C^{LR}_{8G}(x)
\longrightarrow
\sum_{i,k}g^2_{H_i}\,\dfrac{{m'_{F_k}}}{M_{A_{H_i}}^2}\,C_{L,R}^{ski*}\,C_{R,L}^{bki}\,C^{LR}_{8G}(x_{ki})\,,
\end{aligned}
\label{SumRule}
\end{equation}
where $x_{ki}\equiv {m^{\prime\,2}}_{F_k}/M^2_{A_{H_i}}$ and $C_{L,R}^{(s,b)ki}$ are the
couplings among the light quarks $(s,b)$ with $A_{H_i}$ and $F_k$. In
this substitution we take $A_{H_i}$ to be the mass eigenstates of the neutral gauge
bosons. Similarly, we may also generalise Eq.~\eqref{eq:LLnewLight} and
\eqref{eq:LRnewLight}.

Notice that in the case that the masses of the heavy particles span a wide range of values, the generalisation above is only a first approximation: a more precise result follows from a detailed analysis using the results of Sect. \ref{sec:QCD} considering several threshold scales. This further study goes beyond the scope of the present paper and in the following we only consider a single high matching scale.

The factor $m'_{F_k}/M^2_{A_{H_i}}$ can represent a dangerous enhancement of the LR (RL) contribution in Eq.~(\ref{SumRule}), which is unlikely to vanish, barring accidental cancellations. We identify a series of conditions under which this contribution exactly vanishes:
\begin{description}
\item 1)\quad $g_{H_i}=g_{H}$, i.e. there is only one gauge symmetry or all the new gauge coupling constants have the same strength;
\item 2)\quad $m'_{F_k}=m'_{F}$, $M_{A_{H_i}}=M_{A_H}$, i.e. all the exotic fermions and the neutral gauge bosons are degenerate in mass, as in the case of unbroken flavour symmetries; 
\item 3)\quad $\sum_{i}C_{L,R}^{ski*}\,C_{L,R}^{bki}=\sum_{i}C_{L,R}^{ski*}\,C_{R,L}^{bki}=0$, i.e. the couplings are unitary matrices satisfying $C_L=C_R$.
\end{description}
Similar conditions apply also to the contributions of SM down-type quarks. 

These conditions assure the exact cancellation of the NP contributions. However, similarly to the GIM mechanism, if they are only partially satisfied, the terms in Eq.~(\ref{SumRule}) do not cancel each other completely and the NP contributions could still be dangerous.

Furthermore, we remark that condition 2) would correspond to degenerate SM fermions and thus these conditions cannot be satisfied in models which successfully explain the SM fermion mass pattern through the see-saw mechanism illustrated in Sect.~\ref{sec:Context}. In these models, however, the mixings between SM and exotic fermions are small, cf. Eqs.~\eqref{CouplingfF} and \eqref{CouplingFf}. This provides a sufficiently strong suppression to safely neglect these contributions. We further discuss this point in Sect.~\ref{sec:3classes}.

%
%
\section{QCD corrections}
\label{sec:QCD}
\subsection{General structure}
\label{subsec:QCDgeneralstructure}

In order to complete the analysis of the $b\to s\gamma$ decay we have to include QCD
corrections, which play a very important role within the SM, enhancing
the rate by a factor of $2-3$~\cite{MABCC:BSGalpha2}. It originates dominantly from the mixing of {\it charged} current-current operators into
the dipole operators and to a smaller extent from the mixing with QCD-penguin operators.

When the flavour gauge bosons $A_{H_i}$ and the exotic fermions $F_k$ are integrated
out at the high scale $\mu_H$, in addition to the dipole operators discussed at length
in the previous section, also {\it neutral} current-current operators corresponding to
a tree-level $A_{H_i}$ exchange are generated together with ordinary QCD-
and QED-penguin operators.

The contributions of QCD- and QED-penguin operators, arising from diagrams
with $A_{H_i}$ and $F_k$ exchanges, are much less important than within the SM. Indeed,
in the SM context, QCD-penguins do not have to compete with tree-level FCNC 
diagrams, as the latter are absent due to the GIM mechanism. However, 
in the present case there are neutral current-current operators originating from
the tree-level $A_{H_i}$ exchanges that are not suppressed by $\alpha_s(\mu_H)$ and
loop effects, contrary to QCD-penguins. Thus, for all practical purposes the contributions of
QCD-penguins from $A_{H_i}$ and $F_k$ exchanges at the high scale $\mu_H$ can be neglected. Even less important
are effects from QED-penguins. Note that this is in contrast to dipole operators
which cannot be generated at tree-level, but nevertheless mediate the $b\to s\gamma$ decay.

In what follows we will extend the SM RG analysis by considering
the QCD effects of neutral current-current operators at scales $\mu_b\leq \mu\leq\mu_H$
generated by diagrams with the exchange of a single neutral gauge boson $A_H$. The
extension to the case of an arbitrary number of such gauge bosons is straightforward. This
analysis does take into account the mixing under QCD renormalisation of these neutral current-current operators 1) into dipole operators, 2) among themselves, 3) into QCD-penguin operators.
The latter mixing generates contributions to QCD-penguin operators at scales lower than $\mu_H$, even if the initial conditions at $\mu_H$ of the Wilson coefficients of these operators are neglected. Even if this effect is numerically
negligible, we include it for completeness as the mixing of QCD-penguin operators and
dipole operators is taken into account in the SM part.

Before going into details, let us note that the neutral current-current operators do
not mix with the charged current-current operators as their flavour structures
differ from each other. On the other hand, similarly to charged current-current
operators, neutral current-current operators have an impact on dipole operators
and QCD-penguins through RG effects without being themselves affected
by the presence of these two types of operators.

We would like to emphasise that the QCD analysis presented below is also relevant for other processes, such as non-leptonic two-body $B$ decays, and other observables, like $\epsilon'/\epsilon$.

Denoting symbolically charged current-current, QCD-penguin,
dipole, and neutral current-current operators and the corresponding primed operators respectively by
\begin{equation}
Q^{cc}\,,\qquad\quad
Q_P\,,\qquad\quad
Q_D\,,\qquad\quad
Q^{nn}\,,\qquad\quad
Q'_P\,,\qquad\quad
Q'_D\,,\qquad\quad
Q^{nn\,\prime}\,,
\end{equation}
the structure of the one-loop anomalous dimension matrix looks as follows:
\begin{equation}
\begin{tabular}{l|c|c|c|c|c|c|c|l}
\multicolumn{1}{r}{}& \multicolumn{1}{c}{$Q^{cc}$} & \multicolumn{1}{c}{$Q_P$}  & \multicolumn{1}{c}{$Q_D $}	& \multicolumn{1}{c}{$Q^{nn}$}
                                                   & \multicolumn{1}{c}{} 	& \multicolumn{1}{c}{}		& \multicolumn{1}{c}{} \\
\cline{2-5}
$Q^{cc}$ 	& $X_1$ 	& $X_2$		& $X_3$ 	& 0 	&\multicolumn{1}{c}{}	&\multicolumn{1}{c}{}	&\multicolumn{1}{c}{}\\
\cline{2-5}
$Q_P$ 		& 0 		& $X_4$ 	& $X_5$ 	& 0 	&\multicolumn{1}{c}{}	&\multicolumn{1}{c}{}	&\multicolumn{1}{c}{}\\
\cline{2-5}
$Q_D$ 		& 0 		& 0 		& $X_6$ 	& 0 	&\multicolumn{1}{c}{}	&\multicolumn{1}{c}{}	&\multicolumn{1}{c}{}\\
\cline{2-5}
$Q^{nn}$ 	& 0 		& $Y_1$ 	& $Y_2$ 	& $Y_3$ &\multicolumn{1}{c}{}	&\multicolumn{1}{c}{}	&\multicolumn{1}{c}{}\\
\cline{2-8}
\multicolumn{1}{c}{} 	   	&\multicolumn{1}{c}{}&\multicolumn{1}{c}{}&\multicolumn{1}{c}{}&	& $X_4$ & $X_5$ & 0 	& \multicolumn{1}{l}{$Q_P'$}\\
\cline{6-8}
\multicolumn{1}{c}{}  	    	&\multicolumn{1}{c}{}&\multicolumn{1}{c}{}&\multicolumn{1}{c}{}&	& 0     & $X_6$ & 0 	& \multicolumn{1}{l}{$Q_D'$}\\
\cline{6-8}
\multicolumn{1}{c}{}  		&\multicolumn{1}{c}{}&\multicolumn{1}{c}{}&\multicolumn{1}{c}{}&	& $Y_1$	& $Y_2$ & $Y_3$ & \multicolumn{1}{l}{$Q^{nn\,\prime}$}\\
\cline{6-8}
\multicolumn{1}{r}{}		&\multicolumn{1}{r}{}&\multicolumn{1}{r}{}&\multicolumn{1}{r}{}&\multicolumn{1}{r}{}& \multicolumn{1}{c}{$Q_P'$}  & \multicolumn{1}{c}{$Q_D' $} & \multicolumn{1}{c}{$Q^{nn\,\prime}$}
\end{tabular}
\label{AnomalousDimension}
\end{equation}
The non-vanishing entries denoted by $X_i$ are known from the SM analysis \cite{CFRS:BSG,BBL:WeakNLO}. The new
results in the present paper are the entries denoted by $Y_i$, the initial
conditions for $Q_D$, given already in Sect. \ref{sec:analytic}, and the
corresponding conditions for $Q^{nn}$, given at the end of this section. 

The RG analysis of the NP contributions can be performed
independent from the SM one: that is 
\begin{equation}
\begin{aligned}
&C_i^{SM}(\mu_b)=U_{ij}(\mu_b,\, \mu_W)\,C_j^{SM}(\mu_W)\,,\\[2mm]
&C_i^{NP}(\mu_b)=W_{ij}(\mu_b,\,\mu_H)\, C_j^{NP}(\mu_H)\,,\\[2mm]
&C_i^\text{total}(\mu_b)=C_i^{SM}(\mu_b)+C_i^{NP}(\mu_b)\,,
\end{aligned}
\end{equation}
where $U_{ij}$ and $W_{ij}$ are the elements of the RG evolution
matrix. The Wilson coefficients for the primed operators at the high scale are
in general different from the expression above, but the evolution matrices $U$ and $W$ do not change.
The general structure of these matrices in terms of the anomalous dimension
matrices and the QCD-$\beta$-functions is well known \cite{Buras:LH-lectures}
and will not be repeated here. Therefore, we present in the next sections our
results for the matrices $Y_i$ and the initial conditions for $Q^{nn}$ and
$Q^{nn\,\prime}$ operators.

\subsection{Operator basis for neutral current-current operators}

There are $48$ neutral current-current operators generated by the exchange
of $A_H$, all containing the neutral currents $(\ov{s}\,\gamma_\mu\,P_{L,R}\,b)$
and the flavour conserving currents $(\ov{f}\,\gamma_\mu\,P_{L,R}\,f)$ with
$f=u,c,t,d,s,b$. At this scale also the exotic fermions have been already integrated out at $\mu_H$.

The general notation for the $48$ $Q^{nn}$ and $Q^{nn\,\prime}$ operators in
question will be 
\begin{equation}
Q_{1,2}^f(A,B)\,,\qquad \qquad \text{for} \quad A,B=\{L,R\}\,.
\label{QnnGeneral}
\end{equation}
For instance
\begin{equation}
\begin{aligned}
&Q_1^u(L,R)=\left(\ov{s}_\alpha\,\gamma_\mu\,P_L\,b_\beta\right)\left(\ov{u}_\beta\,\gamma^\mu\,P_R\,u_\alpha\right)\,,\\[2mm]
&Q_2^u(L,R)=\left(\ov{s}_\alpha\,\gamma_\mu\,P_L\,b_\alpha)(\ov{u}_\beta\,\gamma^\mu\,P_R\,u_\beta\right)\,,
\end{aligned}
\end{equation}
where $\alpha,\beta=1,2,3$ are colour indices.

Moreover there are 8 additional neutral current-current operators
that have to be included, as they have a different structure, when colour
and flavour structures are considered simultaneously:
\begin{equation}
\begin{aligned}
&\hat{Q}_1^d(A,B)=\left(\ov{s}_\alpha\,\gamma_\mu\,P_A\,d_\beta\right)\left(\ov{d}_\beta\,\gamma^\mu\,P_B\,b_\alpha\right)\,,\\[2mm]
&\hat{Q}_2^d(A,B)=\left(\ov{s}_\alpha\,\gamma_\mu\,P_A\,d_\alpha)(\ov{d}_\beta\,\gamma^\mu\,P_B\,b_\beta\right)\,.
\end{aligned}
\label{QnnhatGeneral}
\end{equation}
In this classification, we refer to $Q^{nn}$ [$Q^{nn\,\prime}$] as those operators
$Q_{1,2}(A,B)$ with $A=L$ [$A=R$] and $B=L,R$ [$B=R,L$].

We point out that the operator basis in Eqs.~(\ref{QnnGeneral}) and
(\ref{QnnhatGeneral}) can be reduced using Fierz transformations. Moreover
some relations can be found between the usual charged current-current
and QCD-penguin operators
\begin{equation}
\begin{aligned}
&Q_1=\left(\ov{s}_\alpha\,\gamma_\mu\,P_L\,c_\beta\right)\left(\ov{c}_\beta\,\gamma^\mu\,P_L\,b_\alpha\right)\,,\\[2mm]
&Q_2=\left(\ov{s}_\alpha\,\gamma_\mu\,P_L\,c_\alpha\right)\left(\ov{c}_\beta\,\gamma^\mu\,P_L\,b_\beta\right)\,,\\[2mm]
&Q_3=\left(\ov{s}_\alpha\,\gamma_\mu\,P_L\,b_\alpha\right)\sum_{q=u,d,s,c,b}\left(\ov{q}_\beta\,\gamma^\mu\,P_L\,q_\beta\right)\,,\\[2mm]
&Q_4=\left(\ov{s}_\alpha\,\gamma_\mu\,P_L\,b_\beta\right)\sum_{q=u,d,s,c,b}\left(\ov{q}_\beta\,\gamma^\mu\,P_L\,q_\alpha\right)\,,\\[2mm]
&Q_5=\left(\ov{s}_\alpha\,\gamma_\mu\,P_L\,b_\alpha\right)\sum_{q=u,d,s,c,b}\left(\ov{q}_\beta\,\gamma^\mu\,P_R\,q_\beta\right)\,,\\[2mm]
&Q_6=\left(\ov{s}_\alpha\,\gamma_\mu\,P_L\,b_\beta\right)\sum_{q=u,d,s,c,b}\left(\ov{q}_\beta\,\gamma^\mu\,P_R\,q_\alpha\right)\,.
\end{aligned}
\end{equation}

Yet, similarly to \cite{BJL:AnatomyEpsilon,BBH:AnatomyEpsilon} we have decided
to work with all operators and not use such relations when performing the RG analysis in order to keep the anomalous dimensions in a
transparent form. As discussed in \cite{BJL:AnatomyEpsilon,BBH:AnatomyEpsilon}
one can  work with linearly dependent operators in the process of RG
analysis without any problems and use such
relations only at the end of the evolution if necessary. However, in the case
of $C_{7\gamma}(\mu_b)$ at LO this is not required. 

In this context we should warn the reader that at NLO level a use of Fierz
relations in the RG analysis could lead  to wrong results within
the NDR scheme unless Fierz evanescent operators are introduced 
\cite{JP:QCDinDeltaS,BMU:2loopQCDGamma}.

\subsection{Anomalous dimension matrices}

We define the LO anomalous dimension matrix through 
\begin{equation}
\hat{\gamma}(\alpha_s)=\dfrac{\alpha_s}{4\pi}\hat{\gamma}^{(0)}\,,
\end{equation}
and give below the results for $\hat{\gamma}^{(0)}$ (dropping the index $(0)$ to simplify the notation). 

The inspection of the one-loop diagrams contributing to the anomalous
dimension matrices shows that only 16 operators have to be considered
in order to find the full matrix. These are
\begin{equation}
\label{ListOperators}
\ba{llll}
Q_{1,2}^u(L,L)\,,&\qquad Q_{1,2}^d(L,L)\,,&\qquad Q_{1,2}^s(L,L)\,,&\qquad \hat{Q}_{1,2}^d(L,L)\,,\\[2mm]
Q_{1,2}^u(L,R)\,,&\qquad Q_{1,2}^d(L,R)\,,&\qquad Q_{1,2}^s(L,R)\,,&\qquad \hat{Q}_{1,2}^d(L,R)\,.
\ea
\end{equation}

The anomalous dimensions for $u$ replaced by $c$ and $t$ are equal to
the one of $Q_{1,2}^u$. The ones of $Q_{1,2}^b$ are equal to the ones
of $Q_{1,2}^s$. The anomalous dimensions of the remaining 28 operators,
namely the primed operators obtained by $L\leftrightarrow R$, are
the same as those of the corresponding unprimed operators 
(see Eq.~(\ref{AnomalousDimension})).

The mixing between the $Q^{nn}$ operators and $Q_D$, that is the matrix
$Y_2$, can be extracted from \cite{CFMRS:BSG,CFRS:BSG} by inspecting the mixing
between QCD-penguin and dipole operators. For the {\it transposed}
matrices we get

\begin{equation}
\hat\gamma_D^T(L,L)=
\begin{tabular}{|c||c|c|c|c|c|c|c|c|}
\hline
&&&&&&&&\\[-4mm]
& $Q_1^u$ & $Q_2^u$ & $Q_1^d$ & $Q_2^d$ & $Q_1^s$ & $Q_2^s$ & $\hat{Q}_1^d$ & $\hat{Q}_2^d$ \\[2mm]
\hline
\hline
&&&&&&&&\\[-4mm]
$Q_{7\gamma}$ & $\dfrac{416}{81}$ & $0$ & $-\dfrac{232}{81}$ & 0 & $-\dfrac{232}{81}$ & $-\dfrac{232}{81}$ & 0 & $-\dfrac{232}{81}$ \\[2mm]
\hline
&&&&&&&&\\[-4mm]
$Q_{8G}$ & $\dfrac{70}{27}$ & $3$ & $\dfrac{70}{27}$ & $3$ & $\dfrac{151}{27}$ & $\dfrac{151}{27}$ & $3$ & $\dfrac{70}{27}$\\[2mm]
\hline
\end{tabular}
\end{equation}
\begin{equation}
\hat\gamma_D^T(L,R)=
\begin{tabular}{|c||c|c|c|c|c|c|c|c|}
\hline
&&&&&&&&\\[-4mm]
& $Q_1^u$ & $Q_2^u$ & $Q_1^d$ & $Q_2^d$ & $Q_1^s$ & $Q_2^s$ & $\hat{Q}_1^d$ & $\hat{Q}_2^d$ \\[2mm]
\hline
\hline
&&&&&&&&\\[-4mm]
$Q_{7\gamma}$ & $-\dfrac{448}{81}$ & $0$ & $\dfrac{200}{81}$ & $0$ & $\dfrac{200}{81}$ & $\dfrac{16}{9}$ & $-\dfrac{80}{3}$ & $-\dfrac{32}{9}$ \\[2mm]
\hline
&&&&&&&&\\[-4mm]
$Q_{8G}$ & $-\dfrac{119}{27}$ & $-3$ & $-\dfrac{119}{27}$ & $-3$ & $-\dfrac{173}{27}$ & $-\dfrac{16}{3}$ & $-4$ & $\dfrac{8}{3}$ \\[2mm]
\hline
\end{tabular}
\end{equation}

The mixing between neutral current-current operators, that is the
matrix $Y_3$, is universally given by two $2\times 2$ matrices:
\begin{equation}
\hat\gamma^{nn}(L,L)=
\begin{tabular}{|c||c|c|}
\hline
&&\\[-4mm]
& $Q_1$ & $Q_2$ \\[2mm]
\hline
\hline
&&\\[-4mm]
$Q_1$ & $-2$ & $6$ \\[2mm]
\hline
&&\\[-4mm]
$Q_2$ & $6$ & $-2$ \\[2mm]
\hline
\end{tabular}\qquad\qquad
\hat\gamma^{nn}(L,R)=
\begin{tabular}{|c||c|c|}
\hline
&&\\[-4mm]
& $Q_1$ & $Q_2$ \\[2mm]
\hline
\hline
&&\\[-4mm]
$Q_1$ & $-16$ & $0$ \\[2mm]
\hline
&&\\[-4mm]
$Q_2$ & $-6$ & $2$ \\[2mm]
\hline
\end{tabular}
\end{equation}
Only operators with the same flavour-conserving structure $(\ov{f}\,\gamma_\mu\,P_{L,R}\,f)$ mix in this way.

Finally, in the case of $Y_1$ there is a universal mixing of $Q^{nn}$
operators into QCD-penguin operators:
\begin{equation}
\hat\gamma_P=
\begin{tabular}{|c||c|c|c|c|}
\hline
&&&&\\[-4mm]
& $Q_3$ & $Q_4$ & $Q_5$ & $Q_6$ \\[2mm]
\hline
\hline
&&&&\\[-4mm]
$Q^{nn}$ & $-\dfrac{2}{9}$ & $\dfrac{2}{3}$ & $-\dfrac{2}{9}$ & $\dfrac{2}{3}$ \\[2mm]
\hline
\end{tabular}
\end{equation}
for
\begin{equation}
Q^{nn}=\left\{Q_1^{u,c,t,d,s,b}(L,L),\, Q_2^{s,b}(L,L),\, \hat{Q}_2^d(L,L),\, Q_1^{u,c,t,d,s,b}(L,R)\right\}
\end{equation}
with no mixing for the remaining operators in Eq.~(\ref{ListOperators}).

\subsection{Initial conditions}

The initial conditions at $\mu_H$ for the dipole operators have already been presented in Sect.~\ref{sec:analytic}: see Eqs.~(\ref{c7}), (\ref{c8}), (\ref{LLnew}), (\ref{LRnew}), (\ref{eq:LLnewLight}) and (\ref{eq:LRnewLight}). Here we give the corresponding initial conditions for the neutral current-current operators.

The initial conditions are determined by integrating out all the
heavy degrees of freedom at $\mu_H$ and matching the result to the effective theory. At LO this matching is trivial. In the following we give the result
for the simplified case in which only one neutral flavour gauge boson $A_H$
contributes to the initial conditions. The generalisation to several
gauge bosons is obvious.

Denoting the vertex $\ov{f}_i\,A_H\,f_j$ as in Eqs.~(\ref{vertexGeneral}) and (\ref{Couplingff}), the general expression for the Wilson
coefficient of the neutral current-current operator
\begin{equation}
Q^f_2(A,B)\equiv (\ov{s}_\alpha\,\gamma_\mu\,P_A\,b_\alpha)(\ov{f}_\beta\,\gamma^\mu\,P_B\,f_\beta)\,,
\end{equation}
in the normalisation of Eq.~(\ref{Heff_at_mu}), is given by
\begin{equation}
\Delta^{AB}C^f_2(\mu_H)=-\dfrac{1}{2}\dfrac{g_H^2}{g_2^2}\dfrac{M_W^2}{M_{A_H}^2}\dfrac{C_A^{sb*}\,C_B^{ff}}{V_{ts}^*\,V_{tb}}\,,
\label{InitialConditionQnn}
\end{equation}
while all the coefficients $\Delta^{AB} C_1$ are zero. Similarly, the Wilson coefficient for the $\hat{Q}^d_{2}(A,B)$ operator is 
\begin{equation}
\Delta^{AB}\hat{C}_2^d(\mu_H)=-\dfrac{1}{2}\dfrac{g_H^2}{g_2^2}\dfrac{M_W^2}{M_{A_H}^2}\dfrac{C_A^{sd*}\,C_B^{bd}}{V_{ts}^*\,V_{tb}}\,,
\label{InitialConditionQnnHat}
\end{equation}
while $\Delta^{AB}\hat{C}_1^d(\mu_H)$ is again zero.

%
%
\section{Phenomenological analysis}
\label{sec:pheno}
\mathversion{bold}
\subsection{The $\ov{B}\to X_s \gamma$  branching ratio}
\mathversion{normal}
\label{sec:Br}

The SM prediction for the $\ov{B}\to X_s\gamma$ branching ratio at
NNLO \cite{MABCC:BSGalpha2,GM:BSGdefinition,MS:BSGdefinition} reads,
\begin{equation}
Br(\ov{B}\to X_s \gamma)=(3.15\pm0.23)\times 10^{-4}\,,
\label{BRSMNNLO}
\end{equation}
and has been calculated for a photon-energy cut-off $E_\gamma>1.6$ GeV in the $\ov{B}$-meson rest frame. This is to be compared
with the current experimental value \cite{PDG2010},
\begin{equation}
Br(\ov{B}\to X_s \gamma)=(3.55\pm 0.24 \pm 0.09)\times 10^{-4}\,,
\label{BRexp}
\end{equation}
for the same energy cut-off $E_\gamma$. In the presence of NP the expression
for the branching ratio is given as follows:
\begin{equation}
Br(\ov{B}\to X_s \gamma) = R \left(|C_{7\gamma}(\mu_b)|^2+|C'_{7\gamma}(\mu_b)|^2+N(E_\gamma)\right)\,,
\label{BRtotal}
\end{equation}
where $R=2.47\times10^{-3}$ and $N(E_\gamma)=(3.6\pm0.6)\times10^{-3}$. The parameter $R$ is simply an overall factor, whose determination is discussed in \cite{GM:BSGdefinition,MS:BSGdefinition}, while $N(E_\gamma)$ is a non-perturbative contribution. Calculating the NP contributions in the LO approximation, but including the
SM contribution at the NNLO level we find for $\mu_b=2.5$ GeV
\begin{equation}
C_{7\gamma}(\mu_b)=C^{SM}_{7\gamma}(\mu_b)+\Delta C_{7\gamma}(\mu_b)
\label{eq:C7efftotal}
\end{equation}
where the central value of $C^{SM}_{7\gamma}(\mu_b)$ at the NNLO, corresponding to the value in Eq.~(\ref{BRSMNNLO}), is
given by \cite{MABCC:BSGalpha2,GM:BSGdefinition,MS:BSGdefinition}
\begin{equation}
C^{SM}_{7\gamma}(\mu_b)=-0.3523
\label{eq:C7effSM}
\end{equation}
and the NP one, calculated by us at the LO, by
\begin{equation}  
\begin{split}
\Delta C_{7\gamma}(\mu_b)=&\quad\kappa_7~\Delta C_{7\gamma}(\mu_H) +\kappa_8~\Delta C_{8G}(\mu_H)+\\[2mm]
			       &+\!\!\!\!\!\sum_{\substack{A=L,R\\f=u,c,t,d,s,b}}\!\!\!\!\! \kappa^{f}_{LA}~\Delta ^{LA} C_2^f(\mu_H)
			        +\!\!\!\sum_{A=L,R}\!\!\!\!\hat{\kappa}^{d}_{LA}~\Delta ^{LA} \hat{C}_2^d(\mu_H).
\end{split}
\label{eq:DeltaC7eff}
\end{equation}
Here $\kappa$'s are the NP magic numbers listed in Tab.~\ref{tab:c7magicnumbers}, calculated taking $\alpha_s(M_Z=91.1876\,\text{GeV})=0.118$. The primed coefficient $C'_{7\gamma}(\mu_b)$ can be obtained from
(\ref{eq:C7efftotal})--(\ref{eq:DeltaC7eff}), by interchanging $L\leftrightarrow R$
and taking the initial conditions of the primed Wilson coefficients. In particular, the NP magic numbers listed in Tab.~\ref{tab:c7magicnumbers} are also valid
for the primed case, as QCD is blind to the fermion chirality\footnote{In
the computation of $C'_{7\gamma}(\mu_b)$ we are neglecting the SM contributions
$C_{7\gamma}^{SM\,\prime}(\mu_b)$ and $C_{8G}^{SM\,\prime}(\mu_b)$, and the NP
contributions $\Delta C'_{7\gamma}(\mu_b)$ and $\Delta C'_{8G}(\mu_b)$, since all of
them are suppressed by $m_s/m_b$ with respect the unprimed ones. Our numerical
analyses confirm that this approximation is consistent with errors below the percent
level.}.

\begin{table}[h!]
\begin{center}
\begin{tabular}{|c||r|r|r|r||r|}
  \hline
  &&&&&\\[-4mm]
  $\mu_H$	 	 & 200 GeV 	& 1 TeV		& 5 TeV		& 10 TeV & $M_Z$\\[1mm]
  \hline\hline
  $\kappa_7$		 & 0.524	& 0.457		& 0.408		& 0.390	 &	 0.566 \\			
  $\kappa_8$		 & 0.118	& 0.125		& 0.129		& 0.130	 & 0.111	\\[1mm]		
  \hline
  &&&&&\\[-4mm]
  $\kappa_{LL}^{u,c}$	 & 0.039	& 0.057		& 0.076		& 0.084	 & 0.030	\\			
  $\kappa_{LL}^{t}$	 &-0.002	&-0.003		&-0.002		&-0.001	 & --	\\			
  $\kappa_{LL}^{d}$	 &-0.040	&-0.057		&-0.072		&-0.079	 & -0.032	\\			
  $\kappa_{LL}^{s,b}$	 & 0.087	& 0.090		& 0.090		& 0.090	 & 0.084	\\			
  $\hat{\kappa}_{LL}^{d}$& 0.128	& 0.147		& 0.163		& 0.168	 & 0.116	\\[1mm]
  \hline             
  &&&&&\\[-4mm]                                                                            	
  $\kappa_{LR}^{u,c}$	 & 0.085	& 0.128	 	& 0.173		& 0.193	 & 0.065	\\			
  $\kappa_{LR}^{t}$	 & 0.004	& 0.012	 	& 0.023		& 0.028	 & --	\\			
  $\kappa_{LR}^{d}$	 &-0.015	&-0.025		&-0.036		&-0.041	 & -0.011	\\			
  $\kappa_{LR}^{s,b}$	 &-0.078	&-0.092	 	&-0.106		&-0.111	 & -0.070	\\			
  $\hat{\kappa}_{LR}^{d}$& 0.473	& 0.665		& 0.865		& 0.953	 & 0.383	\\[1mm]	
  \hline                                                                                         	
\end{tabular}
\caption{The NP magic numbers for $\Delta C_{7\gamma}(\mu_b)$ defined
in Eq.~(\ref{eq:DeltaC7eff}). For completeness, we include in the last column the case of a flavour-violating $Z$. The analogous table for 
$\Delta C_{8G}(\mu_b)$ is provided in Appendix \ref{AppA}.}
\label{tab:c7magicnumbers}
\end{center}
\end{table}

The SM contribution containing NLO and NNLO QCD corrections exhibits a negligible
$\mu_b$-dependence. This is not the case for the NP contribution at LO.
However, we have checked numerically that when the NP contribution enhances the SM value
by $20\%$, the $\mu_b$-dependence in the total branching ratio amounts
to a $3\%$ uncertainty for $\mu_b\in[2.5,\,5]$ GeV. For smaller deviations from the SM prediction 
the uncertainty further decreases. This renders this uncertainty sufficiently small
for our purposes.\\

Before concluding a few observations are in order:
\begin{description}
\item 1)\quad Similarly to the SM, the magic numbers $\kappa_7$ and $\kappa_8$ suppress the initial values $\Delta C_{7\gamma}(\mu_H)$ and $\Delta C_{8G}(\mu_H)$. This is due to the QCD RG evolution running down to $\mu_b$. Furthermore, the suppression of $\Delta C_{7\gamma}(\mu_H)$  increases with $\mu_H$.
 
\item 2)\quad As in the SM, provided $\Delta^{AB}C_2^f(\mu_H)$ and $\Delta^{AB}\hat{C}_2^d(\mu_H)$
  are sufficiently larger than $\Delta C_{7\gamma}(\mu_H)$, the additive QCD corrections stemming from the mixing of the neutral current-current operators
  into the dipole operators are dominant. Furthermore the QCD factors $\kappa_i$ increase in
  most cases with $\mu_H$.
  The most prominent is the coefficient $\hat{\kappa}_{LR}^d$ which
  could even be of $\cO(1)$, but also $\kappa_{LR}^{u,c}$, $\hat{\kappa}^d_{LL}$ and
  $\kappa^{s,b}_{LL,LR}$ are sizable. However, values of the corresponding initial conditions could compensate these coefficients, as in the case of small couplings among SM fermions.
  
\item 3)\quad The SM and NP primed dipole Wilson coefficients, $ C'_i(\mu_W)$ and  $\Delta C'_i(\mu_H)$, are suppressed by $m_s/m_b$ and turn out to be numerically negligible. On the other hand, this suppression is absent in the contributions of the primed neutral current-current operators $Q^{nn\prime}$ and therefore they should be considered in the determination of the branching ratio.
\end{description}

\subsection{Three classes of models}
\label{sec:3classes}

The results for $\kappa_i$ shown in Tab.~\ref{tab:c7magicnumbers} are
model-independent and hold for all models in which the neutral gauge
bosons have flavour violating couplings to fermions as
discussed in Sect.~\ref{sec:Context}.
On the other hand, the initial conditions of the Wilson coefficients are
model-dependent. However, in spite of the large variety of models it is possible to distinguish three main classes, even if
hybrid situations are also possible:
\begin{description}
  \item 1)\quad Models without exotic fermions. This is the case
  of models that do not aim at explaining the SM fermion masses and mixings
  or models that do provide such an explanation but without exotic fermions. For instance, $Z'$ constructions in which the theory is 
  anomaly free without extending the fermion spectrum
  of the SM fall into this class. This is possible when the Green--Schwarz mechanism is implemented in the theory or when the
  generator of the additional $U(1)$ is
  a linear combination of the hypercharge $Y$ and $B-L$, where $B$ is the
  baryon number and $L$ is the lepton number.

  For all models of this class, the new diagrams with  virtual
  exotic fermions do not contribute. On the other hand, the contributions from the exchange
  of the light down-type quarks, the neutral current-current operators
  and the corresponding QCD evolution are still present. As illustrated
  in our numerical analysis of Sect.~\ref{sec:ToyModel}, the light-quark
  contributions turn out to be negligible in the whole parameter space.

\item 2)\quad Models in which the SM fermion-mass patterns are governed
  by exotic fermions through a see-saw mechanism as illustrated in
  Sect.~\ref{sec:Context}. In this case, the couplings of
  the neutral gauge bosons to SM and exotic fermions, described by
  Eqs.~(\ref{Couplingff})--(\ref{CouplingFf}), are suppressed by
  $\sin\theta_{L_k,R_k}$ given in Eqs.~(\ref{FormulaSin1}) and (\ref{FormulaSin2}).
  In the specific limit of $m'_{F_k}\gg m_{f_k}$ and $M_1^D\approx M_2^D\equiv M^D$
  the suppression in $\sin\theta_{L_k,R_k}$ is approximately
  of order $\cO\left(\sqrt{m_{f_k}/m'_{F_k}}\right)$ .

  In order to illustrate the size of the contributions from the new diagrams
  of Sect.~\ref{sec:analytic}, we redefine the couplings $C_{L,R}^{ski}$
  and $C_{L,R}^{bki}$ to exhibit the dependence on the suppressing factor. Without
  loss of generality, we can approximately write
  \begin{equation}
  C_{L,R}^{ski}\simeq \sqrt{\dfrac{m_s}{m'_{F_k}}} \tilde{C}_{L,R}^{ski}\,,\qquad\qquad
  C_{L,R}^{bki}\simeq \sqrt{\dfrac{m_b}{m'_{F_k}}} \tilde{C}_{L,R}^{bki}\,.
  \end{equation}
  As a result, for the case of arbitrary $A_{H_i}$ and $F_k$, the expressions
  on the right-hand side of Eqs.~(\ref{LLnew}) and (\ref{LRnew}) are given by
  \begin{equation}
  \begin{aligned}
    \Delta^{LL}C^{\text{heavy}}_{7\gamma}(\mu_H) &\simeq-\dfrac{1}{6}\,\sum_{i,k}\dfrac{g_{H_i}^2}{g_2^2}\,\dfrac{M_W^2}{M_{A_{H_i}}^2}\,\dfrac{\sqrt{m_s\,m_b}}{m'_{F_k}}\,\dfrac{\tilde{C}_L^{ski*}\,\tilde{C}_L^{bki}}{V_{ts}^*\,V_{tb}}\,\left(C_{8G}^{SM}(x_{ki})+\dfrac{1}{3}\right),\\[2mm]
    \Delta^{LR}C^{\text{heavy}}_{7\gamma}(\mu_H)&\simeq-\dfrac{1}{6}\,\sum_{i,k}\dfrac{g_{H_i}^2}{g_2^2}\,\dfrac{M_W^2}{ M_{A_{H_i}}^2}\,\sqrt{\dfrac{m_s}{m_b}}\,\dfrac{\tilde{C}_L^{ski*}\,\tilde{C}_R^{bki}}{V_{ts}^*\,V_{tb}}\,C^{LR}_{8G}(x_{ki})\,,
  \end{aligned}
  \label{LRnewnew}
  \end{equation}
  and $\Delta^{(LL,LR)}C^{\text{heavy}}_{8G}(\mu_H)=-3\Delta^{(LL,LR)}C^{\text{heavy}}_{7\gamma}(\mu_H)$.
  Similar expressions can be written for the primed contributions. Notice
  that the $m'_{F_k}/m_b$ enhancement is completely removed from the LR (RL)
  Wilson coefficient and is replaced by the suppressing factor $\sqrt{m_s/m_b}$.
  The only dependence on $m'_{F_k}$ is in the loop factor $C^{LR}_{8G}(x_{ik})$.
  On the other hand, an extra inverse power of $m'_{F_k}$ is now present in the LL (RR) contribution, which further suppresses this term.

As we shall explicitly demonstrate in Sect.~\ref{sec:ToyModel}
  these contributions turn out
  to be negligible with respect to those from the neutral current-current operators.
  The same holds for the light-quark contributions. Hence, when
  dealing with this class of models, we simply neglect the contributions of
  exotic fermions and therefore, as in 1), only neutral current-current operators
  and their mixing with dipole operators are relevant.

\item 3)\quad Models with exotic fermions, in which the definition of their couplings to gauge bosons and SM fermions is independent of
  the mechanism of the SM fermion mass generation. For example, this is the case of models in which
  these couplings do not arise from the standard kinetic terms of
  Eq.~(\ref{StandardKineticTerms}). As a result, $\sin\theta_{L_k,R_k}$ can in general be much larger than in
  the previous classes of models.

  In this case, the expressions in Eqs.~(\ref{LLnew}) and (\ref{LRnew}) do not suffer
  from the additional suppressions of Eq.~(\ref{LRnewnew}) and in particular the LR (RL)
  contribution is strongly enhanced by the factor $m'_{F_k}/m_b$. For the
  models of this class, the contribution of exotic fermions is the dominant one
  and all the other contributions can be safely neglected.
\end{description}

We summarise the relevant features of the models in these three classes
in Tab.~\ref{tab:Categories}.

\begin{table}[h!]
\begin{center}
\begin{tabular}{|l||c|c|c|c|}
  \hline
  &&&\\[-4mm]
  Classes of models						& $Q_D^\text{heavy}$ 	& $Q_D^\text{light}$ 	& $Q^{nn}$,  $Q^{nn\prime}$ \\[1mm]
  \hline\hline
  &&&\\[-4mm]	
  1) without exotic fermions  				& Absent				& Negligible		& Dominant \\
  2) with exotic fermions and see-saw			& Negligible			& Negligible		& Dominant \\
  3) with exotic fermions but without see-saw 	& Dominant			& Negligible		& Negligible \\[1mm]
  \hline                                                                                         	
\end{tabular}
\caption{\it Summary of the different classes of models and the
corresponding NP contributions to $b\to s\gamma$: contributions from
the exchange of heavy exotic quarks $Q_D^\text{heavy}$, from the exchange
of SM down-type quarks $Q_D^\text{light}$, and from neutral current-current
operators $Q^{nn}$ and $Q^{nn\prime}$.
}
\label{tab:Categories}
\end{center}
\end{table}

\subsection{Model-independent constraints}

It has been pointed out in \cite{AGM:SignC7gamma,CMW:SignC7gamma,HW:SignC7gamma,
GOS:SignC7gamma,GHM:SignC7gamma,AABGS:SO10SusyGutFCNC} that considering the
measurements of $\ov{B}\to X_s \ell^+\ell^-$ \cite{BaBar:Bsll,Belle:Bsll} the
sign of $C_{7\gamma}(\mu_b)$ in the presence of NP is likely to be
the same as in the SM.
This provides a first model-independent constraint for NP contributions: since
$C^{SM}_{7\gamma}(\mu_b)$ is negative, $\Delta C_{7\gamma}(\mu_b)$ should
also be negative in order to soften the tension between the central values
of the SM prediction and the experimental determination of 
$Br(\overline{B}\rightarrow X_s\gamma)$. This holds not only for contributions from unprimed operators, but also when considering the primed
ones due to the relative suppression of $C'_{7\gamma}(\mu_b)$ with respect to 
$C_{7\gamma}(\mu_b)$, as explicitly confirmed by our numerical analysis.

The sign of the NP contributions is determined by the product of the initial conditions and the QCD magic numbers in
Tab.~\ref{tab:c7magicnumbers}. The signs of the $\kappa_i$ factors are fixed solely by the QCD running, while the sign of the initial conditions 
$\Delta C_i(\mu_H)$ depends on the couplings of the new neutral gauge bosons to the fermions.

In particular, for a model in the first two classes, this constraint translates into the requirement
that the second line on the right-hand side of Eq.~(\ref{eq:DeltaC7eff})
should be negative, namely that:
\begin{equation}
\Delta C_{7\gamma}(\mu_b)\simeq\sum_{\substack{A=L,R\\f=u,c,t,d,s,b}}\kappa^{f}_{LA}~\Delta ^{LA} C_2^f(\mu_H)+
\sum_{A=L,R}\!\!\!\!\hat{\kappa}^{d}_{LA}~\Delta ^{LA} \hat{C}_2^d(\mu_H)<0\,,
\label{Condition1_12}
\end{equation}
 where only the couplings listed in Eq.~(\ref{Couplingff}) are involved. 
 
 On the other
 hand, when a model belongs to the third class, the first line on the right-hand
 side of Eq.~(\ref{eq:DeltaC7eff}) must be negative,
 \begin{equation}
 \Delta C_{7\gamma}(\mu_b)\simeq\kappa_7~\Delta C_{7\gamma}(\mu_H) +\kappa_8~\Delta C_{8G}(\mu_H)<0\,.
 \label{Condition1_3}
 \end{equation} 
  This puts a constraint on the couplings of Eqs.~(\ref{CouplingfF}) and (\ref{CouplingFf}).

  Once the coupling constants are chosen such that the NP contributions have
  the same sign as the SM one, it is possible to further constrain the parameter space
  by requiring that the predicted branching ratio should not exceed the experimental
  bound. From Eq.~(\ref{BRtotal}) we
  find the constraint
  \begin{equation}
  -\Delta C_{7\gamma}(\mu_b)+1.4 \left(\left|\Delta C_{7\gamma}(\mu_b)|^2+|\Delta C'_{7\gamma}(\mu_b)\right|^2\right)\lesssim4.2 (6.1) \times 10^{-2},
  \label{Condition2}
  \end{equation}
  corresponding to the $1\,\sigma$ ($2\,\sigma$) departure from the experimental value, Eq.~\eqref{BRexp}.
  
  It is straightforward to apply these constraints to all models belonging
  to one of the three classes discussed above. It is not
  possible to obtain more insight without specifying a particular model (an analysis will follow in \cite{BCMS:progress}).
  However, in the next section, we provide a simplified representative for each class of models
 and show how the constraints discussed so far apply to each of them.

\subsection{Toy-model examples}
\label{sec:ToyModel}

For each class discussed above, we consider a toy-model in order to justify
the approximations made in the previous sections and to exemplify the application
of the model-independent constraints.

\begin{description}
  \item {\it Classes 1) and 2):}\quad Here, the relevant initial conditions are those
  presented in Eqs.~(\ref{InitialConditionQnn}) and (\ref{InitialConditionQnnHat}).
  The coupling constants entering these expressions are $C_{L,R}^{sb}$, $C_{L,R}^{ff}$,
  $C_{L,R}^{sd}$ and $C_{L,R}^{bd}$, with $f=u,c,t,d,s,b$. To simplify the
  analysis, we assume that all flavour-violating couplings with a
  strange flavour are equal to $C_{FV}^{s}$, the two bottom couplings $C_{L,R}^{bd}$ equal to $C_{FV}^{b}$ and all the flavour conserving ones
  equal to $C_{FC}$. Furthermore, we take
  $V_{ts}=-0.04047$, $V_{tb}=0.999146$ \cite{CKMfitter:2005}, assume that $g_H=g_2$, and consider only one heavy neutral gauge boson. In this way
  we have defined a toy model with only four free parameters: three coupling constants and the mass of the neutral gauge boson.
  
  In Fig.~\ref{fig:C7comparison1} on the left we show the breakdown of 
  $C_{7\gamma}(\mu_b)$ in its different contributions
  as a function of the coupling constants $C_{FV}\equiv C_{FV}^{s}= C_{FV}^{b}=C_{FC}$, for $M_{A_H}=1$ TeV. For completeness, we also show the exotic-quark contributions for $m'_F=10$ TeV. As expected from the discussion
  in Sec.~\ref{sec:3classes}, all the NP contributions apart from the neutral 
  current-current are negligible, justifying our approximations in the previous section.
  We remark that $C^\prime_{7\gamma}(\mu_b)$ almost coincides with the neutral current-current contribution (red line).
  In Fig.~\ref{fig:C7comparison1} on the right we show the value of the coupling constants $C_{FV}$, as
  a function of $M_{A_H}$, for which the bound in Eq.~(\ref{Condition2}) is saturated at the 
  $1 \sigma$ and $2\sigma$ level. As we can see, for small values of $M_{A_H}$ the
  couplings are constrained to small values.
  
    \begin{figure}[h!]
\centering
\includegraphics[]{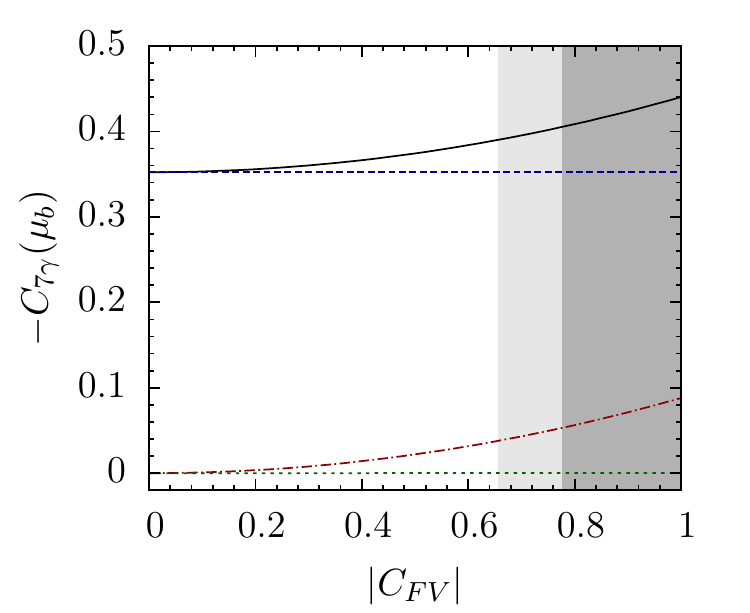}
\hspace*{-10pt}
\includegraphics[]{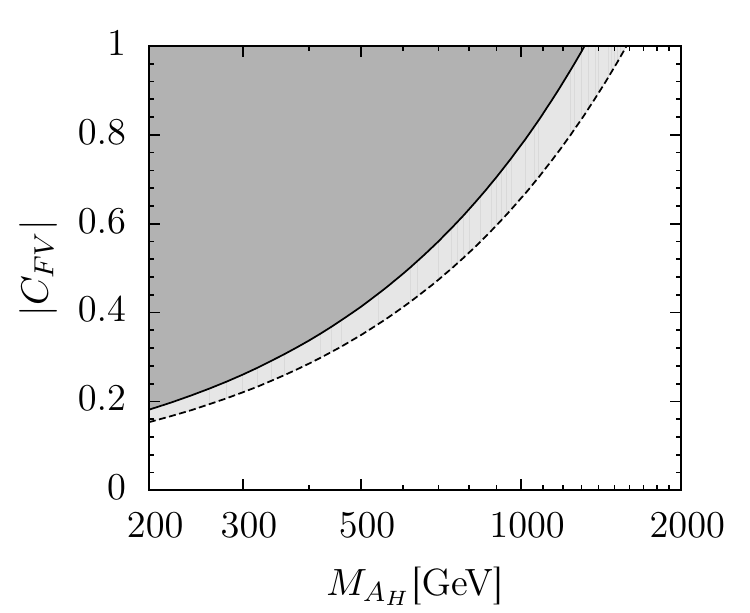}
\caption{\it On the left, the different contributions to $C_{7\gamma}(\mu_b)$
(solid line) are plotted as functions of $C_{FV}$: in blue the SM contribution, in red
the neutral current-current contribution, in green the overlapping contributions from exotic and SM down-type quarks. The shadowed region is excluded
imposing the bound in Eq.~(\ref{Condition2}) at $1\sigma$ (lighter) and $2\sigma$ (darker). On the right, we show the value of
$C_{FV}$ as a function of $M_{A_H}$ for which the bound in  Eq.~(\ref{Condition2})
is saturated. Again the shadowed region represents the excluded values.
\label{fig:C7comparison1}}
\end{figure}

In Fig.~\ref{fig:Constraint12} we separately present the implementation of the two model-independent
constraints, for $M_{A_H}=1$ TeV. On the left, the constraint on the sign of $\Delta C_{7\gamma}(\mu_b)$ 
mostly reduces the parameter space to cases in which the coupling constants 
$C_{FV}^{s}$ and $C_{FV}^{b}$
have opposite signs. In Fig.~\ref{fig:Constraint12}
on the right, the bound from the branching ratio also applies, further constraining the parameter space.

\begin{figure}[h!]
\centering
\includegraphics[]{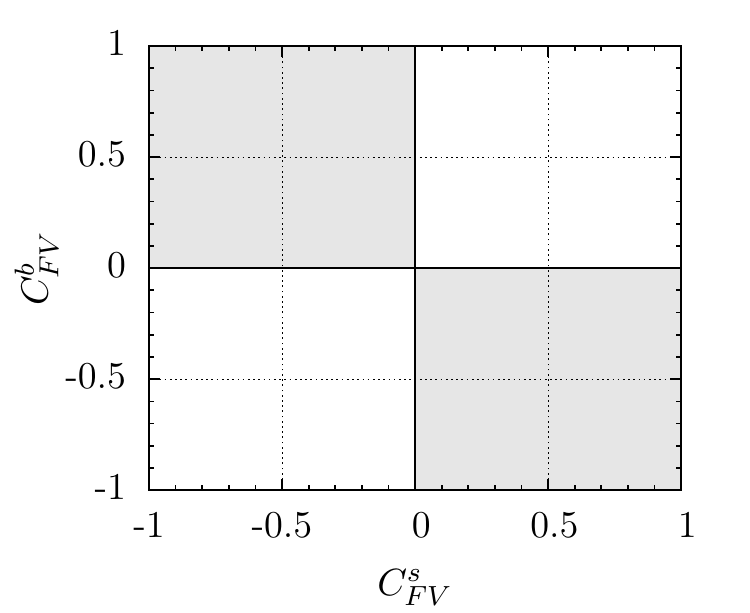}
\hspace*{-10pt}
\includegraphics[]{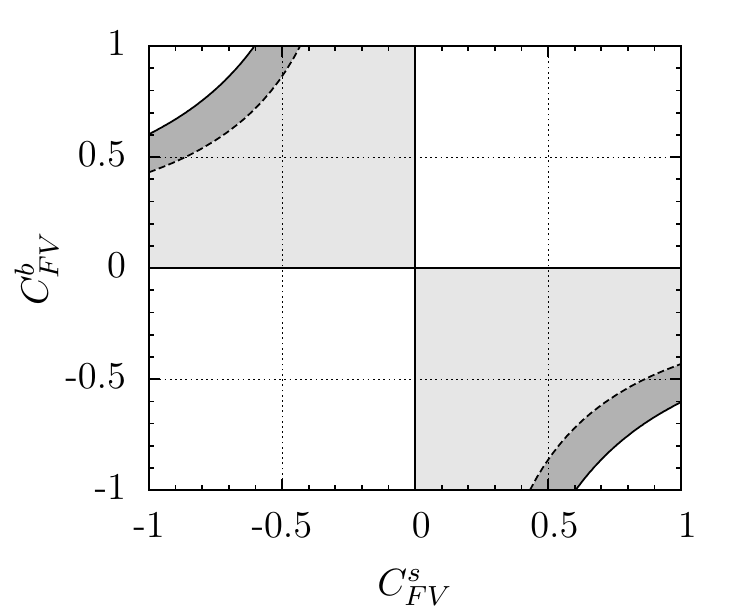}
\caption{\it On the left (right), the points satisfying $\Delta C_{7\gamma}(\mu_b)<0$
(Eq.~(\ref{Condition2})) in the plane $C_{FV}^{s}$ vs. $C_{FV}^{b}(\equiv C_{FC})$.
The shadowed regions now represent the values for which the bounds are passed: lighter (darker) intensity refers to $-4.2 (6.1)\times10^{-4}$. 
\label{fig:Constraint12}}
\end{figure}

\item {\it Class 3):}\quad We consider here the case with only one heavy neutral
  gauge boson and one exotic fermion. The relevant initial conditions are those in
  Eqs.~(\ref{LLnew}) and (\ref{LRnew}). The coupling constants which enter these
  expressions are $C_{L,R}^{sF}$ and $C_{L,R}^{bF}$. Fixing $g_H=g_2$, $M_{A_H}=1$ TeV
  and $m'_F=10$ TeV, and identifying the coupling constants $C_{FV}\equiv C_{L,R}^{sF}=C_{L,R}^{bF}$, we illustrate in Fig.~\ref{fig:C7comparison2} on the left that the dominant NP contributions stems solely from exotic quarks. For a complete comparison, we also plot the contributions from light quarks and neutral current-current operators, adopting the same conventions as for Fig.~\ref{fig:C7comparison1}.

  In Fig.~\ref{fig:C7comparison2} on the right we show the value for the coupling
  constants $C_{FV}$ as function of $M_{A_H}$ for which the $1\sigma$ and $2\sigma$ bounds in Eq.~(\ref{Condition2})
  are saturated. As we can see the constraint on the couplings is very strong and only
  a small part of the shown parameter space survives even for large values of $M_{A_H}$.
  
  \begin{figure}[h!]
\centering
\includegraphics[]{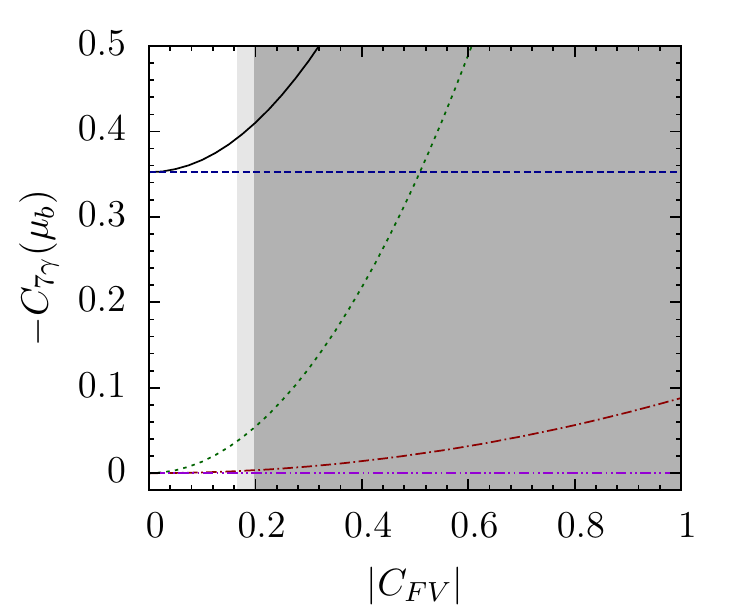}
\hspace*{-10pt}
\includegraphics[]{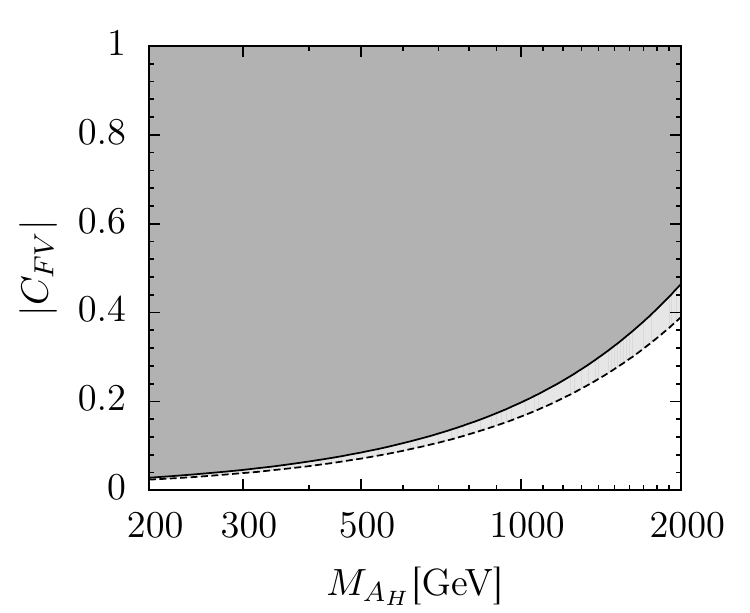}
\caption{\it A similar description to Fig.~\ref{fig:C7comparison1} applies. Here, however, the green (purple) line refers to the exotic-quark (SM-quark) contributions.
\label{fig:C7comparison2}}
\end{figure}

  In Fig.~\ref{fig:Constraint3} on the left (right) we show how the couplings are
  constrained by the expression in Eq.~(\ref{Condition1_3}) (Eq.~(\ref{Condition2})):
  a negative value for $\Delta C_{7\gamma}(\mu_b)$ is only recovered when both
  $C_L^{sF}$ and $C_{L,R}^{bF}$ have the same sign, while the lower bound
  in Eq.~(\ref{Condition2}) provides a very strong constraint on the parameters of the model.

\begin{figure}[h!]
\centering
\includegraphics[]{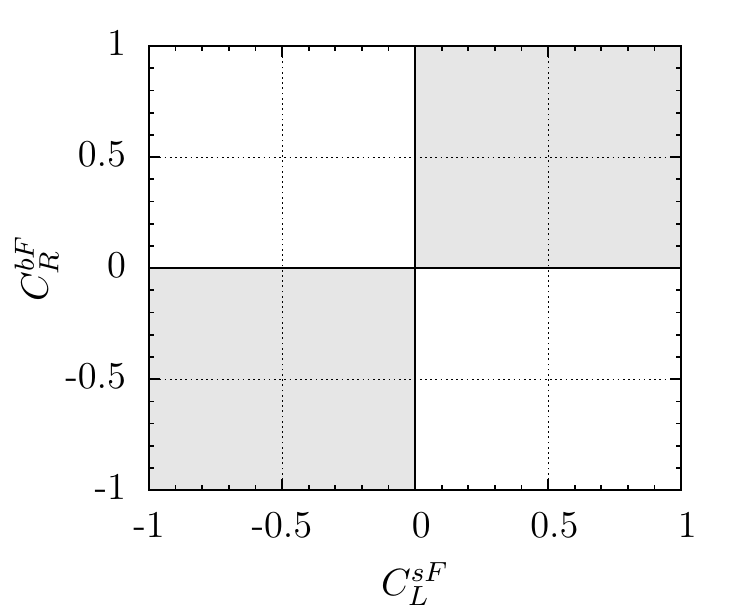}
\hspace*{-10pt}
\includegraphics[]{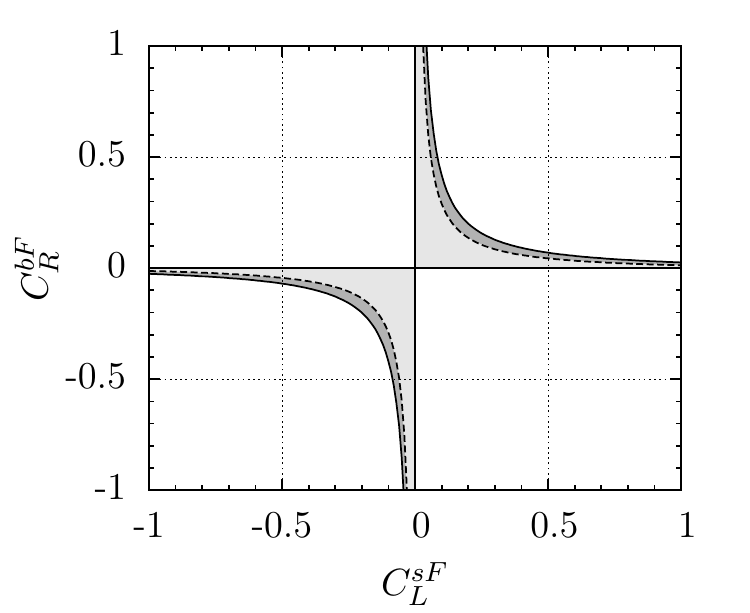}
\caption{\it On the left (right), the points satisfying $\Delta C_{7\gamma}(\mu_b)<0$
(Eq.~(\ref{Condition2})) in the plane $C_L^{sF}$ vs. $C_R^{bF}$, for $C_L^{bF}\in[-1,1]$. The shadowed regions are defined in Fig.~\ref{fig:Constraint12}. 
\label{fig:Constraint3}}
\end{figure}
\end{description}

%
%
\section{Conclusions}
\label{sec:concl}

Extensions of the SM in which the gauge group is enlarged by additional symmetries are attractive, because they provide an explanation of the flavour puzzle and predictions testable at colliders. In these models heavy exotic fermions are usually introduced in order to cancel possible anomalies and to justify the observed SM fermion spectrum through the see-saw mechanism. On the other hand, the presence of new heavy neutral gauge bosons and exotic fermions in principle translates into NP contributions to FCNC processes. 

In this paper, we have pointed out two new contributions to the $\ov{B}\to X_s\gamma$ decay that arise in such models. The first is generated through one-loop diagrams mediated by neutral gauge bosons and exotic down-type quarks. The relevance of this contribution depends on the strength of the left-handed and right-handed flavour violating couplings of the neutral gauge bosons to the SM and the exotic quarks. Analogous effects can be present in other processes like $\mu\to e\gamma$ and $t\to c\gamma$ as well as flavour conserving observables like $(g-2)_\mu$ and EDMs, in which dipole operators play the dominant role.

The second contribution is due to the presence of neutral current-current operators, mediated by neutral gauge bosons, and arises only if the neutral gauge bosons have flavour violating couplings to the SM quarks. Through the QCD mixing with the magnetic dipole operator $Q_{7\gamma}$, these new neutral current-current operators contribute to $\ov{B}\to X_s\gamma$. To our knowledge these QCD effects have been calculated here for the first time. Furthermore, our QCD analysis of the mixing among neutral current-current, QCD penguin and gluonic dipole operators could also be relevant for other processes, such as non-leptonic two-body $B$ decays, and other observables, like $\epsilon'/\epsilon$.

Beside these new contributions, we have also considered the contributions arising from one-loop diagrams with the exchange of neutral gauge bosons and SM down-type quarks, that have been only partially analysed in the literature.

We have studied the impact of all these contributions in a model-independent approach and summarised the resulting constraints in Eqs.~(\ref{Condition1_12})--(\ref{Condition2}). We have presented these expressions in such a manner that in order to test a specific model, it is sufficient to specify only the values of the couplings of the neutral gauge bosons to SM and exotic quarks and their masses. In particular, it is not necessary to repeat the QCD analysis.

A detailed application of this study on a concrete NP scenario is in progress. Here, without entering into details of a particular model, we have described three representative classes of models and discussed the relevance of these contributions. For models in the first class, the SM spectrum is enriched only by the neutral gauge bosons, but no exotic quarks are present. In this case the contributions from the exotic quarks are absent, but those from the neutral current-current operators turn out to have a potentially observable effect on the branching ratio of $\ov{B}\to X_s\gamma$. The value of the masses of the neutral gauge bosons and the strength of their flavour violating couplings to SM fermions determine the relevance of this effect.

The second class accounts for models in which the SM fermion masses $m_f$ are explained through the see-saw mechanism with heavy exotic fermions of masses $m'_F$. In this case, the couplings of neutral gauge bosons to SM and exotic fermions are suppressed by terms $m_f/m'_F$. Therefore, the contributions to $Br(\ov{B}\to X_s\gamma)$ from the exchange of exotic quarks turn out to be negligible with respect to those from the QCD mixing of the neutral current-current and magnetic dipole operators.

The models in the third class are characterised by the presence of heavy exotic fermions, which either provide the SM fermion masses through a different mechanism than the see-saw or do not participate at all in the explanation of the SM flavour spectrum. In this case, no suppression occurs in the couplings of neutral gauge bosons to SM and exotic fermions and the contributions to $Br(\ov{B}\to X_s\gamma)$ from exotic quarks are enhanced by terms $m'_F/m_b$, where $m_b$ is the mass of the bottom quark. These contributions dominate over all the others.

For all the models in the three classes, the contributions from SM down-type quarks turn out to be subdominant in the whole parameter space.

Our analysis shows once more how FCNC processes can put constraints on beyond-SM constructions even before the discovery of new particles in high-energy processes.

%
%

\section*{Acknowledgments}
We thank Joachim Brod and Robert Ziegler for useful comments on the preliminary version of the paper and Mikolaj Misiak, Gerhard Buchalla, and Paride Paradisi for interesting discussions. This work was supported by ERC Advanced Grant ``FLAVOUR'' (267104).

\appendix
%
%

\mathversion{bold}
\section{The $\ov{B}\to X_s\, g$ decay}
\label{AppA}
\mathversion{normal} 

Similarly to Eq.~\eqref{eq:C7efftotal} we also evolve $\Delta C_{8G}$ down to
$\mu_b=2.5$ GeV to obtain
\begin{equation}  
\begin{split}
\Delta C_{8\gamma}(\mu_b)=&\quad\rho_7~\Delta C_{7\gamma}(\mu_H) +\rho_8~\Delta C_{8G}(\mu_H)+\\[2mm]
			       &+\!\!\!\!\!\sum_{\substack{A=L,R\\f=u,c,t,d,s,b}}\!\!\!\!\! \rho^{f}_{LA}~\Delta ^{LA} C_2^f(\mu_H)
			        +\!\!\!\sum_{A=L,R}\!\!\!\!\hat{\rho}^{d}_{LA}~\Delta ^{LA} \hat{C}_2^d(\mu_H),
\end{split}
\label{eq:DeltaC8eff}
\end{equation}
with the NP magic numbers $\rho_i$ listed in Tab.~\ref{tab:c8magicnumbers}.

\begin{table}[h!]
\begin{center}
\begin{tabular}{|c||r|r|r|r||r|}
  \hline
  &&&&&\\[-4mm]
  $\mu_H$			 & 200 GeV 	& 1 TeV		& 5 TeV		& 10 TeV & $M_Z$ \\[1mm]
  \hline\hline
  &&&&&\\[-4mm]
  $\rho_7$			 & 0		& 0		& 0		& 0	& 0	\\			
  $\rho_8$			 & 0.568	& 0.504		& 0.456		& 0.439	 & 0.607	\\[1mm]	
  \hline                                         
  &&&&&\\[-4mm]        
  $\rho_{LL}^{u,c}$		 &-0.124	&-0.138		&-0.147		&-0.150		& -0.115\\			
  $\rho_{LL}^{t}$	 	 &-0.015	&-0.033		&-0.046		&-0.050	 &	--\\			
  $\rho_{LL}^{d}$		 &-0.124	&-0.138		&-0.147		&-0.150	 &  -0.115	\\			
  $\rho_{LL}^{s,b}$		 &-0.243	&-0.279		&-0.307	 	&-0.318  &  -0.222	\\			
  $\hat{\rho}_{LL}^{d}$		 &-0.119	&-0.141		&-0.160		&-0.168	 & -0.107	\\[1mm]
  \hline                               
  &&&&&\\[-4mm]                  
  $\rho_{LR}^{u,c}$		 & 0.184	& 0.229		& 0.270		& 0.287	&  0.160	\\			
  $\rho_{LR}^{t}$		 & 0.015	& 0.037 	& 0.055		& 0.062	 &--	\\			
  $\rho_{LR}^{d}$		 & 0.184	& 0.229		& 0.270		& 0.287	 &  0.160	\\			
  $\rho_{LR}^{s,b}$		 & 0.311	& 0.382		& 0.447		& 0.474	 &  0.272	\\			
  $\hat{\rho}_{LR}^{d}$		 &-0.064	&-0.052		&-0.034		&-0.025	 &  -0.067	\\[1mm]			
  \hline                                                                                         	
\end{tabular}
\caption{The NP magic numbers for $\Delta C_{8G}(\mu_b)$ in Eq.~\eqref{eq:DeltaC8eff}. For completeness, we include in the last column the case of a flavour-violating $Z$.
\label{tab:c8magicnumbers}}
\end{center}
\end{table}

%
%

\providecommand{\href}[2]{#2}\begingroup\raggedright\endgroup

\end{document}